# Low Emittance Positron Beam Generation:
# A Comparison Between Photo-production and Electro-Production

Alberto Bacci[1], Francesco Broggi[1], Vittoria Petrillo[1,2], Luca Serafini[1]

[1]INFN-Sezione di Milano, Dip. Scienze Fisiche Via Celoria 16 20133 Milano, Italy and
LASA Lab. Via F.lli Cervi 201 Segrate, Italy
[2] Univ. degli Studi di Milano Via Celoria 16 20133 Milano, Italy

## Abstract

Positron beams, both polarized and non-polarized, low and high energy have a wide range of applications and many methods can be used to produce them. In this paper we analyse the two main methods used to produce positrons for accelerator applications, the "traditional" electro-production and the more recent "photo-production".

After having determined the best target material, the positron production yield is determined for various target thickness and primary beam energy. Production efficiency up to about 7.6% and 200% are achievable for photo-production and electro-production respectively. Then the quality, i.e. the emittance and the normalized emittance of the produced beam is evaluated; the proposed "Gruber" correction for calculating the normalized emittance is considered too.

The brightness of the obtained beams is calculated, relating it to the intrinsic temporal lengthening of the positron beam. The time structure of the primary beam is preserved for long bunches down to the ps scale.

A very preliminary evaluation of the energy deposition in the target and its effect is done. The polarization is not take, so far, into account.

# 1    INTRODUCTION

Positron beams are widely used in many research fields. In high energy machines like ILC [1] or CLIC [2] both polarized and unpolarized positron beams will be used to achieve fundamental physics and precision physical measurements like Higgs boson physics, parton distribution, studies on pion and kaon structure, or test of the Standard Model (charge conjugation violation, etc.). Positron beams can be used for the muon beam generation for the future muon collider by the electron-positron annihilation process [3, 4].

For energies up to a few hundred keV, their use has a wide field of application [5] like surface magnetization of materials [6] or high resolution and accuracy investigation on material inner structural defects [7], or differentiate between different kinds of atoms, as a matter of fact depending on the element, the positron electron annihilation leads to a different gamma spectrum. Positron beams are also used for the study of the Bose-Einstein condensate [8], for anti-matter research [9] and new energy sources.

In the case of collider physics, the luminosity of the colliding is given by

$$\mathcal{L} = \frac{N_1 N_2 f \, N_b}{4\pi\sigma_x\sigma_y} \tag{1}$$

where:  $N_1$ and $N_2$ are the number of particles in the bunch 1 and 2

$f$   if the repetition frequency

$N_b$  in the number of colliding bunches

$\sigma_x \, \sigma_y$ are the transverse dimensions of the beams (supposing equal for the two beams)

It is clear that the intensity (Number of particles) and quality of the beam (emittance that is related to the dimension) are important parameters.

The ways of positron production are [10]:
- The using of positron from β decay (e.g. from $^{22}$Na). The positron spectrum is continuous up to 540 keV. This production method is used in low energy measurements, as previously reported.
- The "classical" two step scheme or electro-production, where an electron beam (from an accelerator) hits a target. The bremsstrahlung radiation in the target then produce electron and positron pairs.
- The photo-production, where a photon beam (from an undulator or Compton backscattered source) hits a target inducing the pair production, in this case this is a one step process.
- The photo-production from laser-electron [11], or laser-solid scattering [12]
- Channeling of electron or radiation in crystalline target [13]
- Other "exotic" method like the electron-plasma interaction or the pairs creation in vacuum [10,14].

One of the main problem in generating high positron flux is the damage induced by the energy deposition and the consequent rise in temperature in the target, inducing mechanical stress; moreover a high repetition rate of the primary pulse can induce shock wave in the target, changing its characteristics, etc.

Another characteristic of the positron source is the polarization of the positrons, that can be transferred by the polarization of the primary beam.

Preliminary analysis of photo-production of positrons has been previously done [15], where the feasibility of producing positron beams at LI2FE (SPARC) facility has been investigated for electron beams of 150-200 MeV energy and 4 MeV photon energy.

Starting from these results in this paper a more detailed study has been carried out, by examining the efficiency of positron generation by electro and photo-production for different primary energies and target materials by using the FLUKA Monte Carlo code [16,17].

Then the target dimensions (1, 2, 5, 10, 12 and 15 mm thickness) and photon energy (4, 10, 20 MeV) effects are evaluated for the Tungsten. A comparison with the electro-production (30, 90, 150 and 900 MeV) electron beam energy is done.

The positron beam characteristics, energy spectrum, emittance and brightness are evaluated relating them to the primary beam characteristics.

We focus only to the production efficiency and on the beam quality, not (for the moment) on the positron polarization, or possible damages of the target.

## 2    PAIR PRODUCTION CROSS SECTION

The cross section for pair production is [18, 19]

$$\sigma_{pp} = \alpha r_e^2 Z^2 P(\varepsilon, Z) \tag{2}$$

where

$\alpha$ : fine structure constant $= 1/137$

$r_e$ : classical electron radius $= 2.818 \, 10^{-15}$ m

$Z$ : Atomic number of the target

The function P($\varepsilon$,Z) depends on the target material and the photon energy through the parameter $\varepsilon = h\nu/m_ec^2$, the photon energy in electron rest mass unit with:

h : Planck constant

$\nu$ : photon frequency

$m_e$ : electron mass

c : velocity of light in vacuum

Pair production can occur in the nuclear electric field or in the atomic electron electric field. Depending on the energy of the incident photon a screening effect of the electric field of the nucleus, by the K shell electrons, must be taken into account.

$$P(\varepsilon, Z) = \begin{cases} \dfrac{28}{9}\ln(2\varepsilon) - \dfrac{218}{27} & \text{for } 1 << \varepsilon << 1/(\alpha Z^{1/3}) \quad \text{Nuclear Field, Low photon energy, no screening} & \text{(a)} \\[2em] \dfrac{28}{9}\ln\dfrac{183}{Z^{\frac{1}{3}}} - \dfrac{2}{27} & \text{for } \varepsilon >> 1/(\alpha Z^{1/3}) \qquad \text{Nuclear fileld, High photon energy, complete screening} & \text{(b)} \\[2em] \dfrac{28}{9}\ln(2\varepsilon) - \dfrac{218}{27} - 1.027 & \text{Nuclear field, outside the above limits but } \varepsilon > 4 & \text{(c)} \\[2em] \dfrac{1}{Z}\left(\dfrac{28}{9}\ln(2\varepsilon) - 11.3\right) & \text{Electron field and } \varepsilon > 4 & \text{(d)} \end{cases} \qquad (3)$$

For the energy considered ($4 < E_\gamma < 20$ MeV the limit of $\varepsilon$ are $7.83 < \varepsilon < 39.14$ while the limit $1/(\alpha Z^{1/3})$ spans from 45.67 (for light element like Cobalt) to 30.35 for Uranium.

So the cross section expressions that best fits our problem are (a) and (c).

## 3   FLUKA GEOMETRY AND PARAMETERS

The primary beam is a monochromatic photon or electron beam with 0.5 mm radius, no divergence, moving along the z axis towards the positive direction, starting at 5 cm from the target (at point 0,0,-5.0). The target border starts at (0,0,0).

The target is a cylinder with a diameter of 4 mm. In the study of the effects of the primary beam dimensions, in case of 10 mm beam radius, the target diameter is 40 mm.

The beam is coaxial to the target. The transport cut-offs for electrons and photons are set at 10 keV,

## 4   FLUKA SIMULATION RESULTS

Just to show the geometry of the problem in fig. 1a, 1b, 1c and 1d the fluence of photons, electrons, positrons and neutrons for 10 MeV photons on 1 mm Tungsten target are shown. The y coordinate extends from -30 cm to 30 cm. Upstream the target there is vacuum while aside and downstream of it there is air.

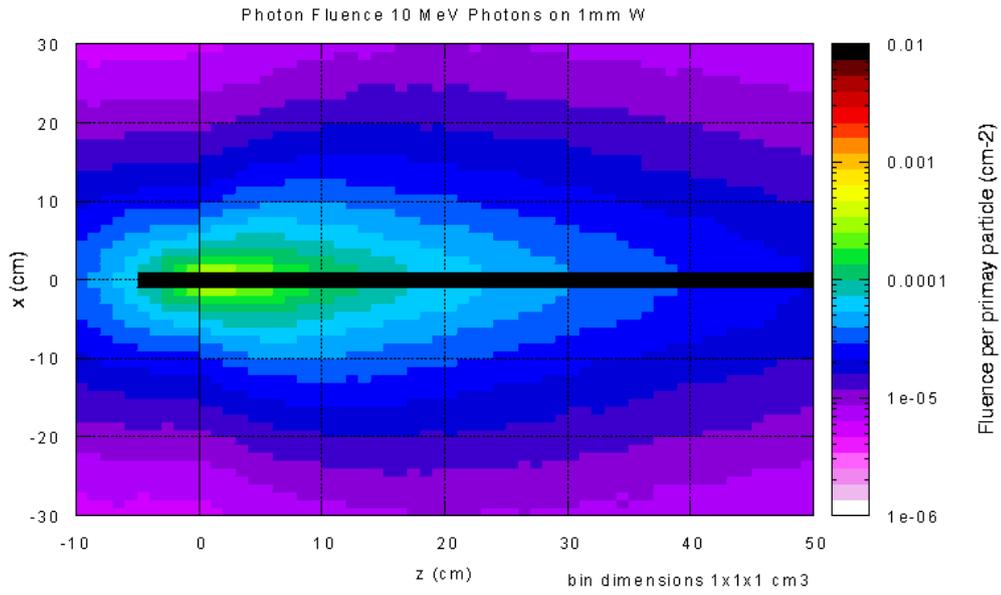

Figure 1a: Photon fluence for 10 MeV photon on 1 mm Tungsten. The primary beam starts at (0,0,-5.0), and moves towards the positive z.

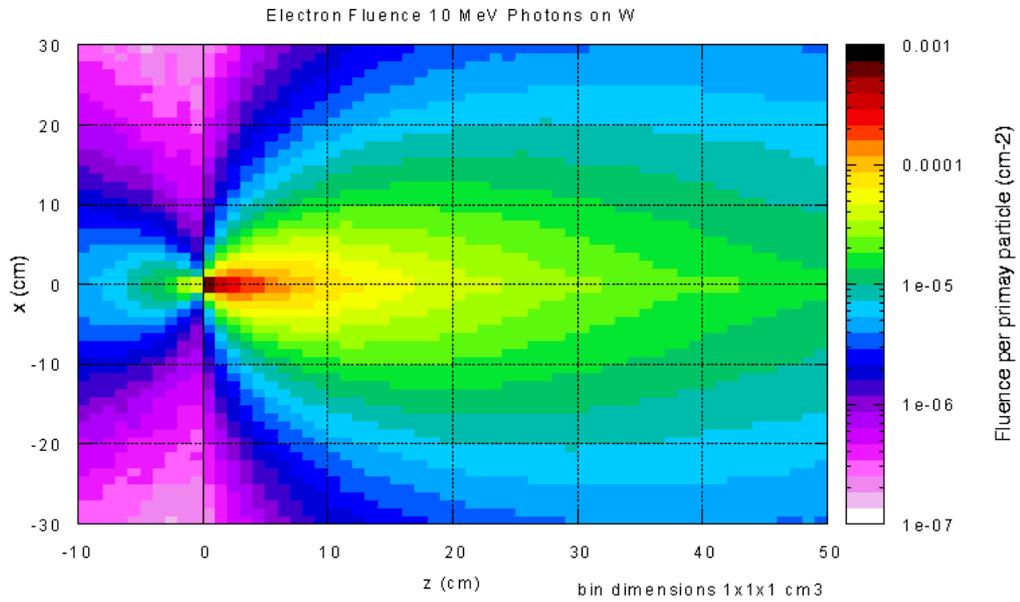

Figure 1b: Electron fluence for 10 MeV photon on 1 mm Tungsten. The primary beam starts at (0,0,-5.0), and moves towards the positive z.

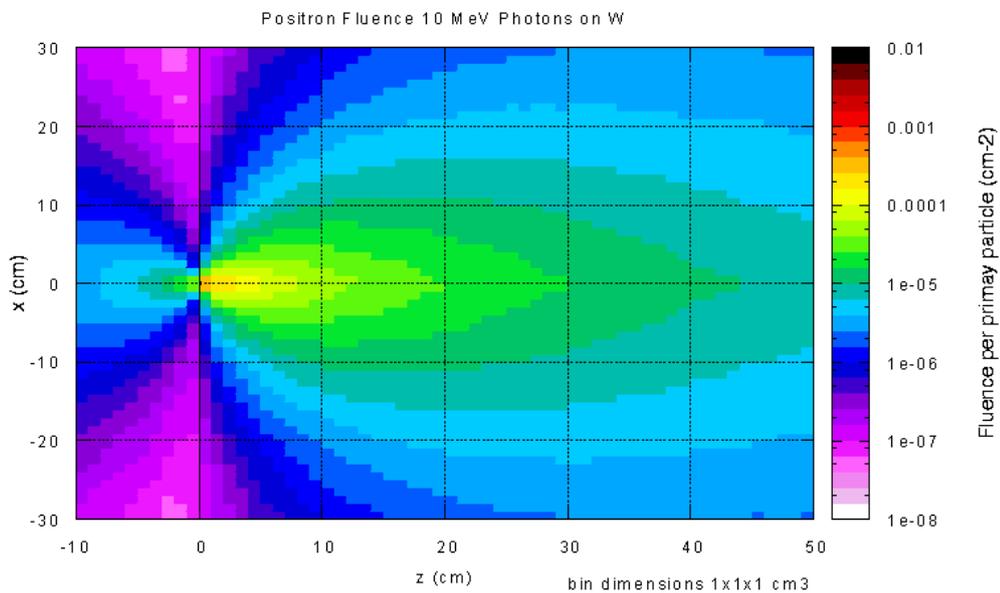

Figure 1c:  Positron fluence for 10 MeV photon on 1 mm Tungsten. The primary beam starts at (0,0,-5.0), and moves towards the positive z.

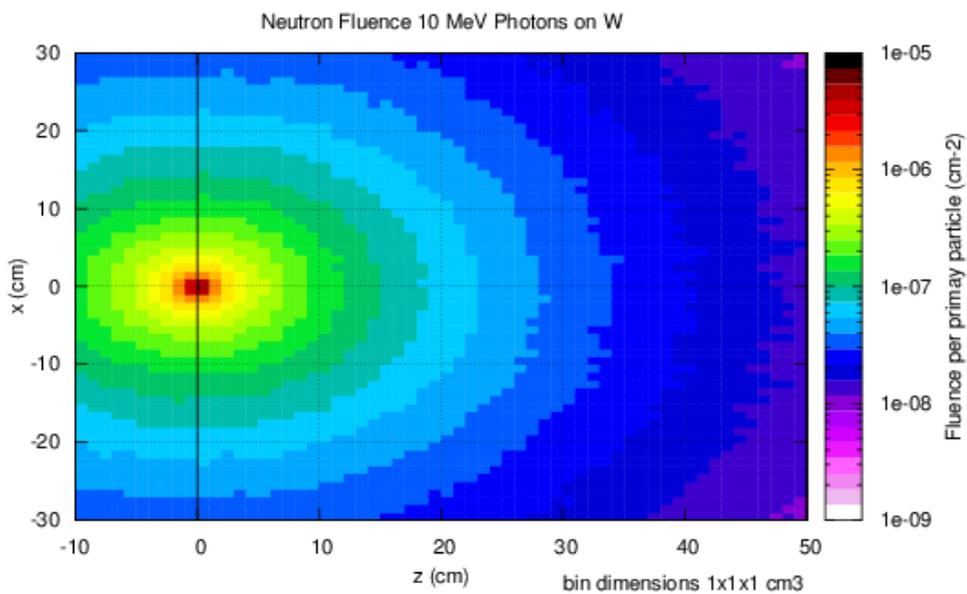

Figure 1d: Neutron fluence for 10 MeV photon on 1 mm Tungsten. The primary beam starts at (0,0,-5.0), and moves towards the positive z.

## 4.1 Target Material

The efficiency of positron production has been investigated for different materials.

The results are obtained with 10 different runs of $10^6$ particles each in case of 10 MeV primary photons hitting a 1 mm thick target.

In fig.2 the number of positrons produced per primary photon in the forward direction for different target materials is shown and compared with a calculation using the cross sections reported above. Cross section (d), the one related to the positron production in the electron field, is neglected because it is 4 orders of magnitude lower than the others.

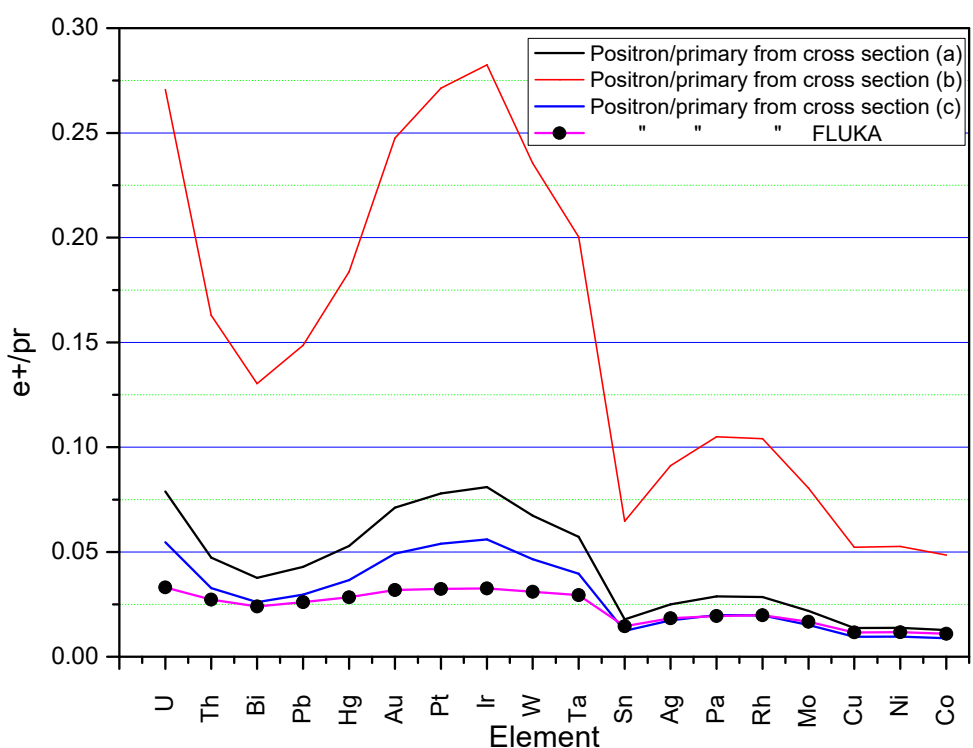

Figure 2: Number of positrons/primary photon in the forward direction for different
materials by 10 MeV photons on 1mm thick target.
Lines: expected as from the theoretical cross sections
Line+symbol: FLUKA results

As expected from the values of ε the cross section expressions that better fits the problem is the (a) and (c) of fig.2. As from the plots in figs. 1 electrons neutrons and photons are produced too.

In fig. 3 the production of these secondary particles is summarized, the data refer to 10 MeV primary photons on 5 mm thick target.

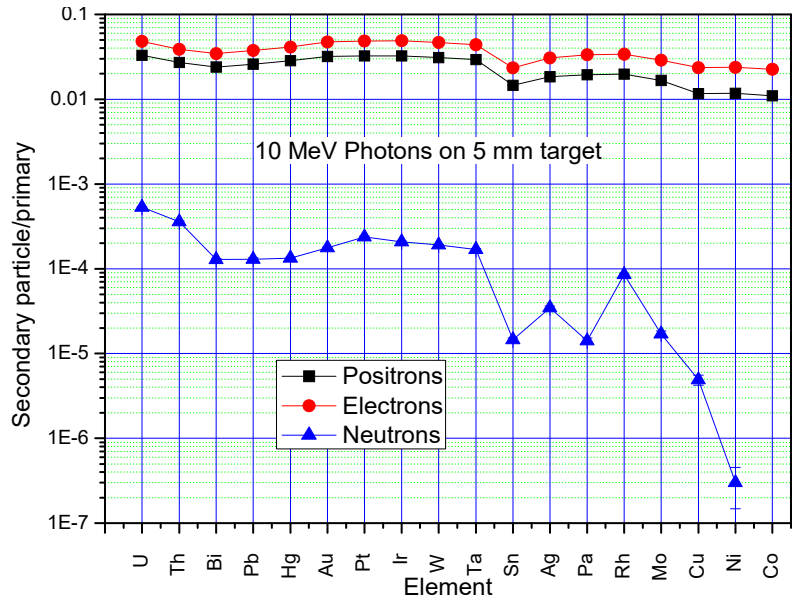

Figure 3: Number of secondary/primary photon in the forward direction for different target materials by 10 MeV photons primary beam, on 5 mm target.

The number of positrons produced is slightly less than the electron ones because electron are produced by Compton scattering in addition to pair production and because of the annihilation occurring in the material, as a matter of fact the spectrum of the photon from the target shows a conversion peak at 511 keV, as shown in fig 4.

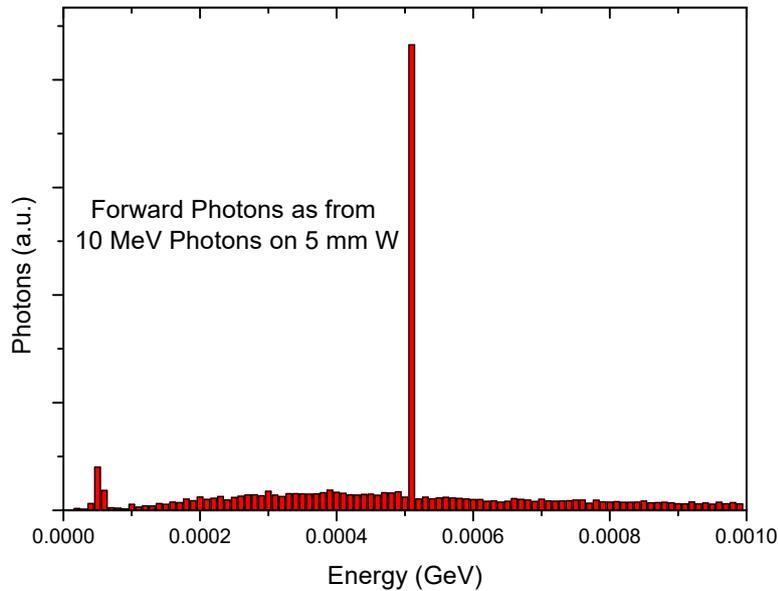

Figure 4: Spectrum of the photons from the target showing the annihilation peak at 511 keV.

## 4.2   Target Geometry

The production of positrons has been investigated for different Tungsten target thickness, both in case of photon and electrons primaries (fig. 5a and 5b respectively).

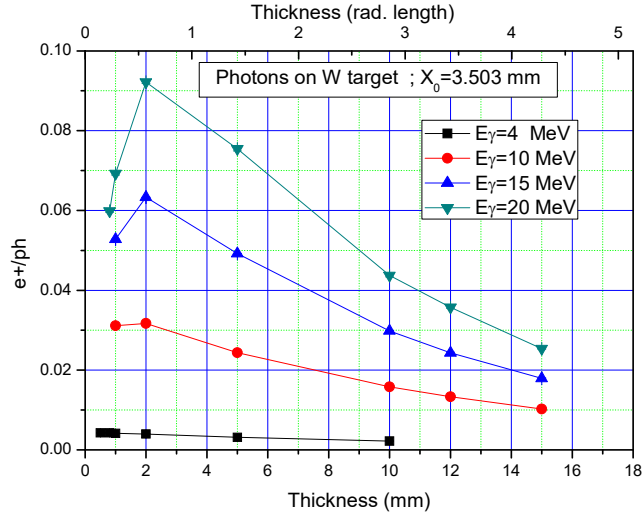

Figure 5a: Number of positrons/primary photon in the forward direction for different Tungsten thickness in mm (bottom scale) or radiation length (top scale), and primary photon energy.

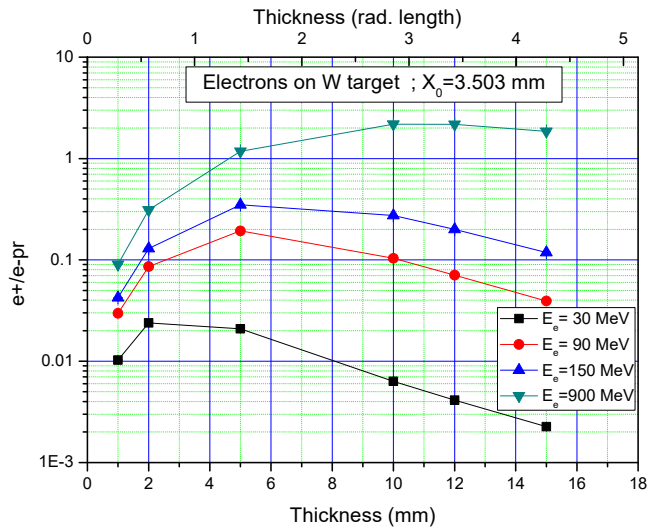

Figure 5b: Number of positrons/primary electron in the forward direction for different Tungsten thickness in mm (bottom scale) or radiation length (top scale), and primary electron energy.

The plots show that the positron yield increase with the primary beam energy (because of the cross section increase).

The best thickness in case of photon primaries is about 2 mm corresponding to $0.6X_0$ ($X_0$= radiation length = 3.5 mm for W); in case of electrons primaries (fig.6b) the best thickness increases with the energy of the primaries, being about 3-4 times the radiation length (or more depending on the electron energy [20]), this because the two stages process of the electro-production is replaced by a single stage process in case of photo-production. Fig. 6 shows the fluencies of the positron inside the target for the 20 MeV photo-production and 900 MeV electro-production in case of thick target.

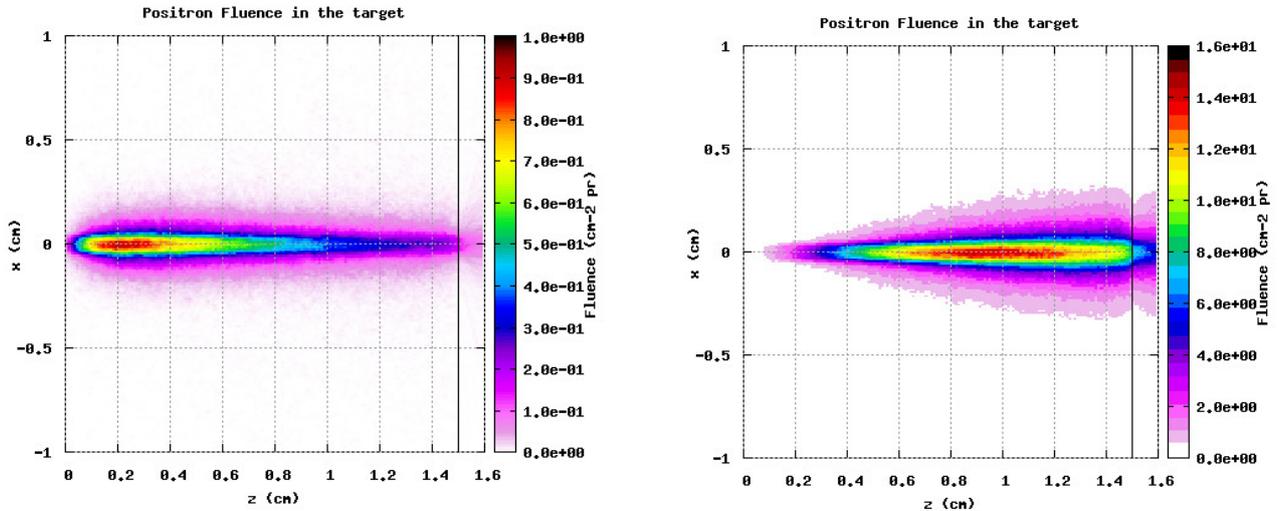

Figure 6: Positron fluencies inside the target for the 20 MeV photo-production (left) and 900 MeV electro-production (right).
The primary beam hits the target (15 mm thickness) at z=0.

Fig. 7 shows the 1D projection of the same fluencies of fig.6.

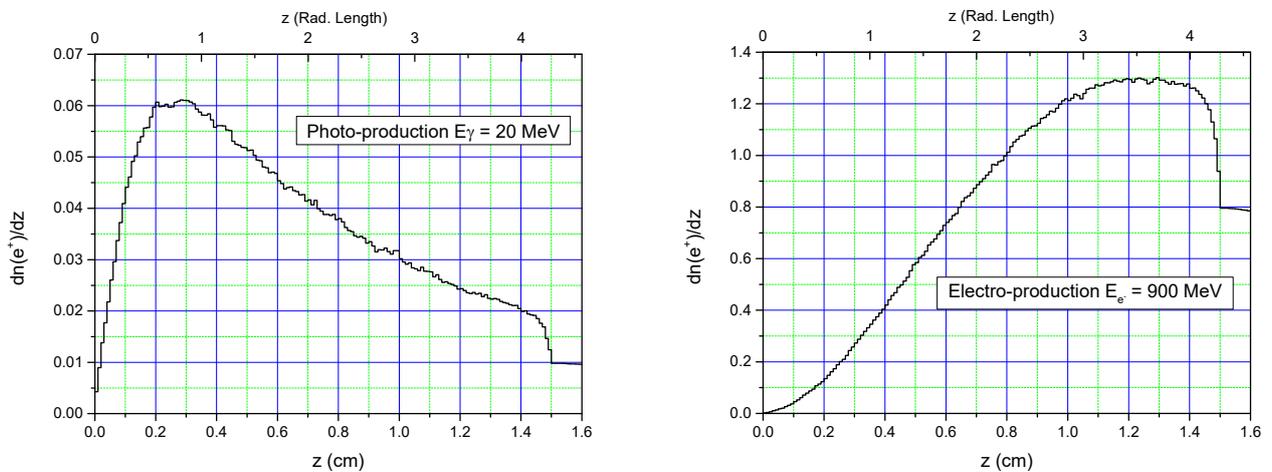

Figure 7: 1D projection of the positron fluencies inside the target for the photo-production (left) and electro-production (right).
The primary beam hits the target (15 mm thickness) at z=0.

Fig. 8 shows the positron spot dimension at the exit of the target for 20 MeV photo-production (at left) and 900 MeV electro-production (right).

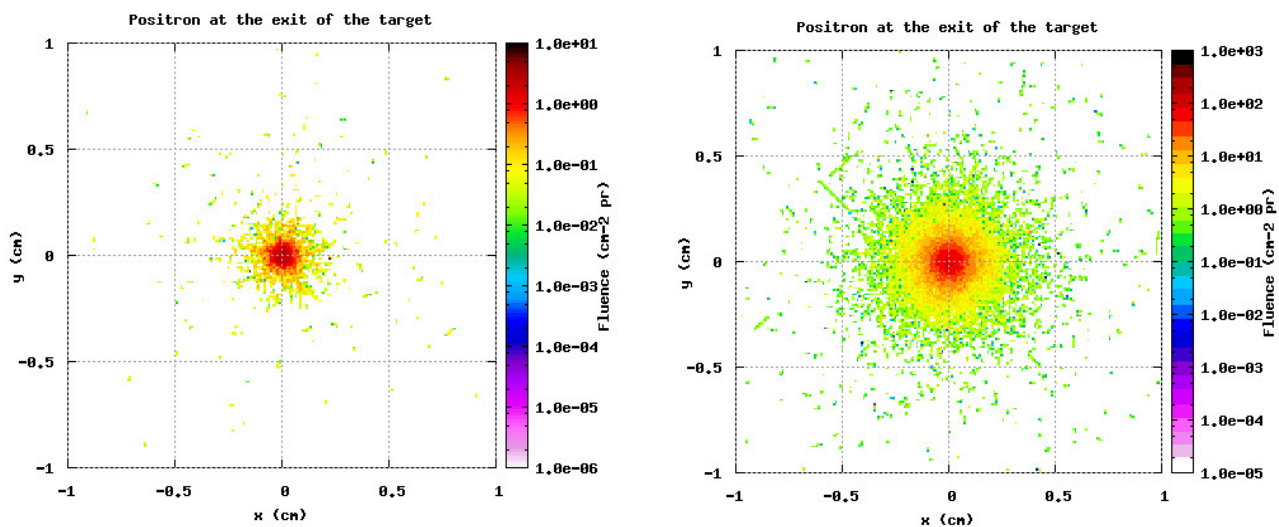

Figure 8: Positron spot at the exit of the 15 mm tungsten target for the 20 MeV photo-production (left) and 900 MeV electro-production (right).

The reduced radiation length can reduce the energy deposition in the target, but this smaller volume can lead to higher peak in temperature so a careful thermal analysis is necessary to evaluate the thermal effects of the energy deposition.

At 4 MeV photon energy the geometry of the target seems unessential for positrons.

The production data reported in the previous figures refer to the total production of positrons, however the emitted positrons have a wide energy spectrum (see next section) and angular distribution.

The particles useful for accelerator applications and transport are the ones emitted in the forward direction. In figs. 9 the total positrons emitted by photon and electron primaries on a tungsten target of 1mm (fig.9a), 2mm (fig.9b), 5mm (fig.9c), 10 mm (fig.9d), 12 mm (fig.9e), and 15 mm (fig.9f) thickness and the ones emitted in a forward cone of different apertures are shown.

In the plots the blue curves refer to the electron production with different primary energy (to be read on the top horizontal scale), while the red curves refer to the photon production (whose primary energy is red on the bottom scale). The symbols refer to the total production (squares), to positrons emitted in a cone of 10° (±5°) aperture (circles), and in a cone of 0.6 aperture (triangles).

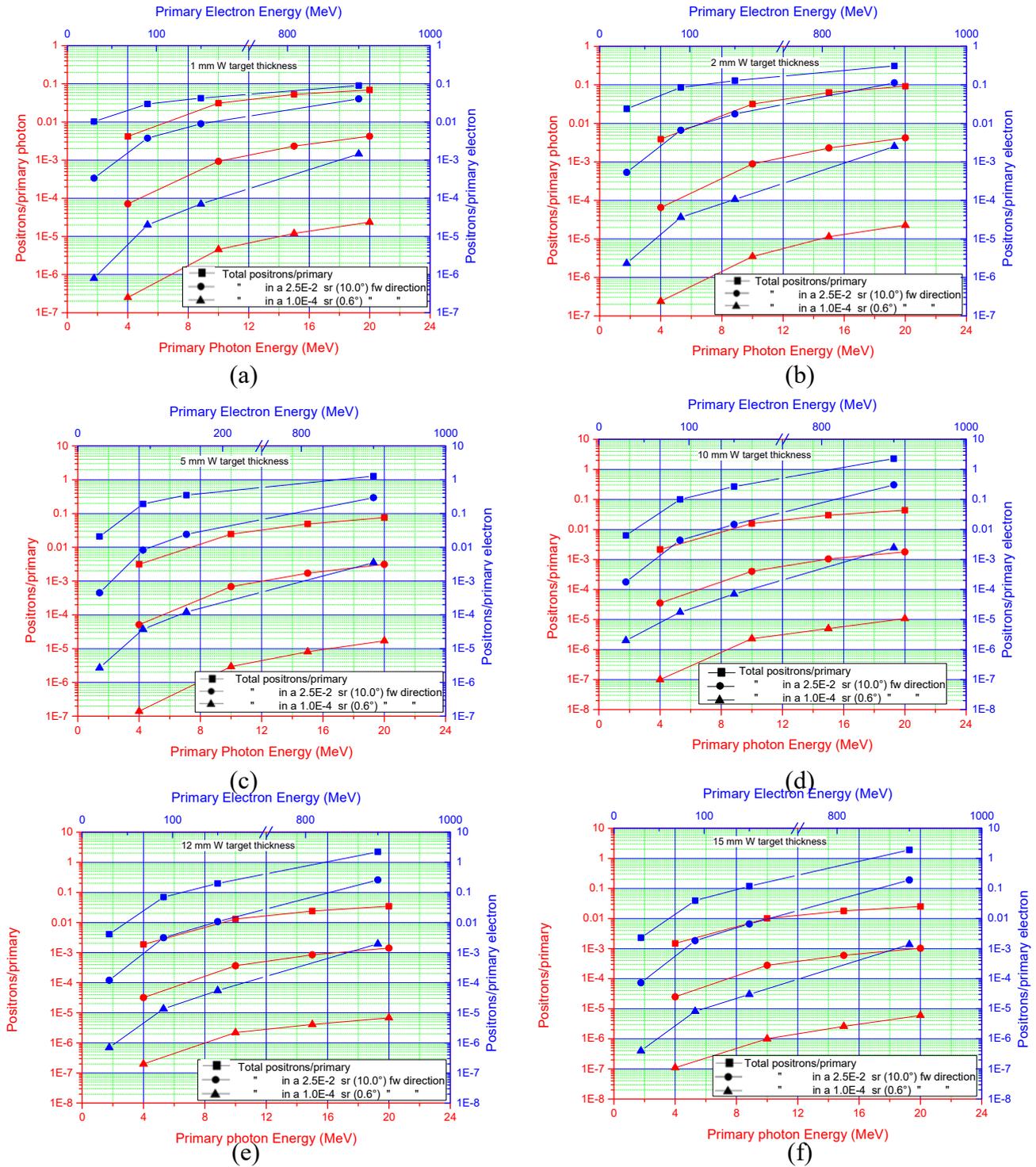

Figure 9: Number of positrons/primary photon (bottom scale)/electron (top scale) in the forward direction for 1mm (a), 2mm(b), 5mm (c), 10 mm(d), 12mm (e) and 15mm (f) tungsten thickness (square symbols), in a cone 10° wide (circles) and 0.6° wide (triangles).

By comparing the plots in figs. 9 we can see that the ratio between the positrons emitted under an angle to the total forward emitted increases as the primary beam energy increases. The positron production by electrons seems more efficient (of about a factor ten), but this depends only on the primary beam energy.

The following tables summarize the data.

In tab. 1a, the total positron produced per primary beam particle, the positrons produced in a cone of 10° and 0.6° aperture together with the percentage error are reported for tungsten target thickness of 1 mm, 2 mm, and 5 mm.

Tab. 1b, is the same as tab. 1a referred at target thickness of 10 mm, 12 mm, and 15 mm.

Table 1a: Positron total yield and forward produced/total positrons for target thickness of 1 mm, 2 mm, and 5 mm.

| Target Thickness | | d = 1 mm | | | d = 2 mm | | | d = 5 mm | | |
|---|---|---|---|---|---|---|---|---|---|---|
| Primary Beam | Primary Energy (MeV) | Total e+ ± err% | e+ in a 10° cone /total | e+ in a 0.6° cone /total | Total e+ ± err% | e+ in a 10° cone /total | e+ in a 0.6° cone /total | Total e+ ± err% | e+ in a 10° cone /total | e+ in a 0.6° cone /total |
| Photons | 4 | 4.16E-03 ± 0.18 | 1.66E-02 | 5.29E-05 | 3.89E-03 ± 0.15 | 1.69E-02 | 6.16E-05 | 3.15E-03 ± 0.13 | 1.69E-02 | 5.09E-05 |
| | 10 | 3.11E-02 ± 0.05 | 2.98E-02 | 1.46E-04 | 3.18E-02 ± 0.04 | 2.78E-02 | 1.10E-04 | 2.46E-02 ± 0.04 | 2.78E-02 | 1.20E-04 |
| | 15 | 5.28E-02 ± 0.06 | 4.42E-02 | 2.28E-04 | 6.33E-02 ± 0.03 | 3.63E-02 | 1.79E-04 | 4.92E-02 ± 0.05 | 3.48E-02 | 1.64E-04 |
| | 20 | 6.91E-02 ± 0.04 | 6.09E-02 | 3.43E-04 | 9.24E-02 ± 0.05 | 4.57E-02 | 2.46E-04 | 7.56E-02 ± 0.04 | 4.13E-02 | 2.25E-04 |
| Electrons | 30 | 1.02E-02 ± 0.32 | 3.30E-02 | 7.81E-05 | 2.39E-02 ± 0.23 | 2.24E-02 | 9.63E-05 | 2.09E-02 ± 0.23 | 2.15E-02 | 1.29E-04 |
| | 90 | 2.97E-02 ± 0.17 | 1.28E-01 | 6.75E-04 | 8.59E-02 ± 0.13 | 7.73E-02 | 4.25E-04 | 1.93E-01 ± 0.08 | 4.28E-02 | 1.94E-04 |
| | 150 | 4.23E-02 ± 0.12 | 2.13E-01 | 1.68E-03 | 1.29E-01 ± 0.05 | 1.36E-01 | 8.26E-04 | 3.52E-01 ± 0.07 | 6.89E-02 | 3.45E-04 |
| | 900 | 9.03E-02 ± 0.07 | 4.49E-01 | 1.61E-02 | 3.12E-01 ± 0.06 | 3.64E-01 | 8.10E-03 | 1.28E-00 ± 0.02 | 2.30E-01 | 2.76E-03 |

Table 1b: Positron total yield and forward produced/total positrons for target thickness of 10 mm, 12 mm, and 15 mm.

| Target Thickness | | d = 10 mm | | | d = 12 mm | | | d = 15 mm | | |
|---|---|---|---|---|---|---|---|---|---|---|
| Primary Beam | Primary Energy (MeV) | Total e⁺ ± err% | e⁺ in a 10° cone/ total | e⁺ in a 0.6° cone /total | Total e⁺ ± err% | e⁺ in a 10° cone /total | e⁺ in a 0.6° cone /total | Total e⁺ ± err% | e⁺ in a 10° cone /total | e⁺ in a 0.6° cone /total |
| Photons | 4 | 2.15E-03 ± 0.57 | 1.65E-02 | 4.65E-05 | 1.86E-03 ± 0.62 | 1.73E-02 | 1.08E-04 | 1.49E-03 ± 0.27 | 1.67E-02 | 7.04E-05 |
| | 10 | 1.58E-02 ± 0.19 | 2.78E-02 | 1.45E-04 | 1.31E-02 ± 0.16 | 2.54E-02 | 1.52E-04 | 1.01E-02 ± 0.25 | 2.87E-02 | 1.01E-04 |
| | 15 | 2.98E-02 ± 0.11 | 3.47E-02 | 1.68E-04 | 2.42E-02 ± 0.16 | 3.45E-02 | 2.06E-04 | 1.78E-02 ± 0.22 | 3.33E-02 | 1.12E-04 |
| | 20 | 4.38E-02 ± 0.18 | 4.07E-02 | 2.47E-04 | 3.49E-02 ± 0.19 | 4.03E-02 | 1.95E-04 | 2.50E-03 ± 0.21 | 4.10E-02 | 2.40E-04 |
| Electrons | 30 | 6.34E-03 ± 0.35 | 2.81E-02 | 3.16E-04 | 4.11E-03 ± 0.50 | 2.93E-02 | 1.70E-04 | 2.29E-03 ± 0.77 | 3.18E-02 | 1.75E-04 |
| | 90 | 1.04E-01 ± 0.09 | 4.22E-02 | 1.69E-04 | 6.97E-02 ± 0.09 | 4.47E-02 | 1.94E-04 | 3.94E-02 ± 0.11 | 4.65E-02 | 2.08E-04 |
| | 150 | 2.74E-01 ± 0.07 | 5.36E-02 | 2.57E-04 | 1.98E-01 ± 0.22 | 5.38E-02 | 2.78E-04 | 1.18E-01 ± 0.25 | 5.61E-02 | 2.55E-04 |
| | 900 | 2.27E-00 ± 0.02 | 1.35E-01 | 1.10E-03 | 2.24E-00 ± 0.08 | 1.16E-01 | 8.74E-04 | 1.91E-00 ± 0.06 | 9.95E-02 | 7.11E-04 |

### 4.3  Positron Spectra

In fig. 10 the energy spectra (integrated over the full solid angle) of the forward produced positrons is shown for the photon (left column) and electron production (right column) from tungsten target. The energy bin is 0.2 MeV wide for the 4 MeV primary photons and 0.5 MeV in all the other cases.

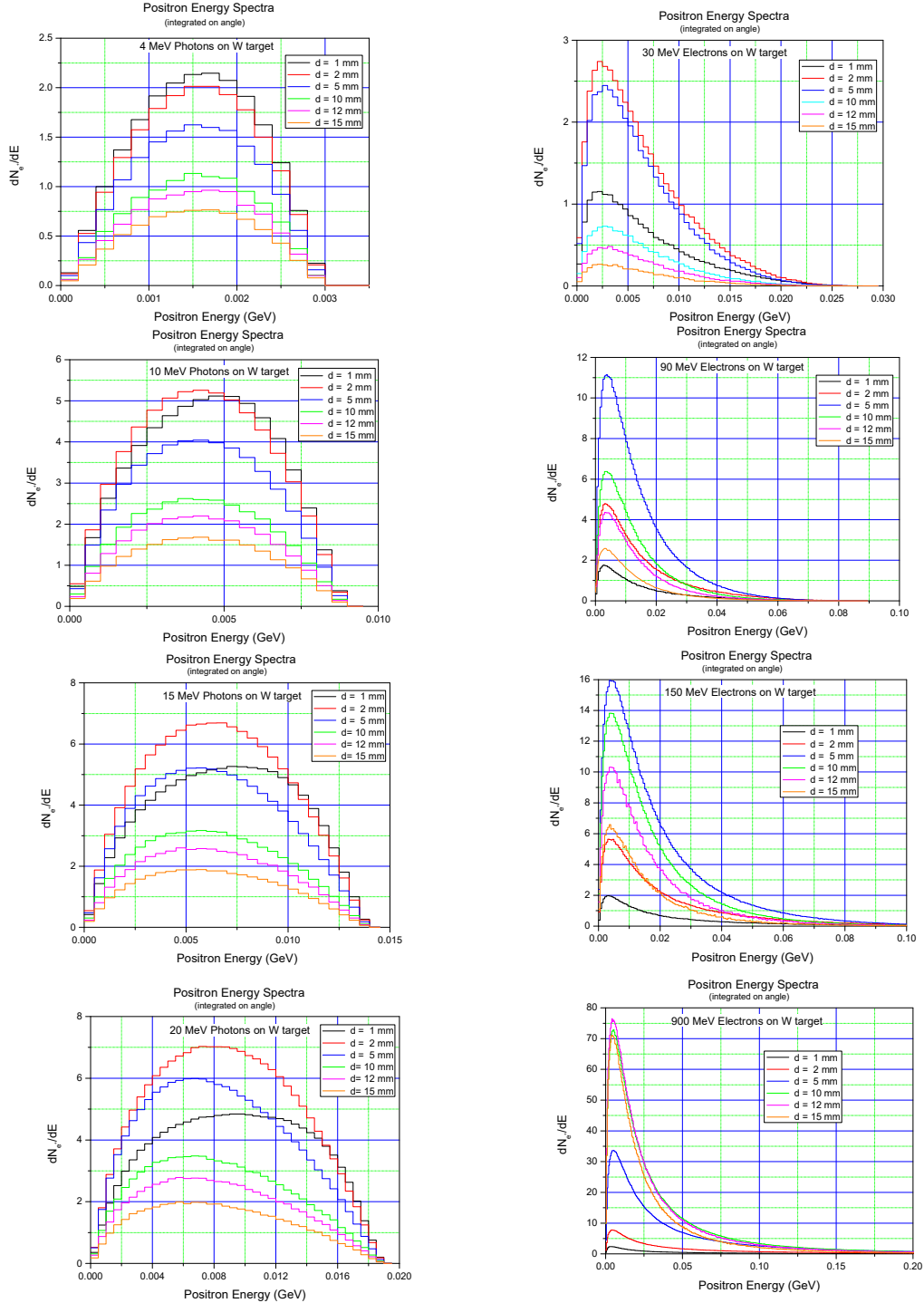

Figure 10: Positron spectra in the forward direction (on the whole solid angle).

The positron spectra from photo-production are more symmetric, with the maximum at about half of the energy of the primary beam. In the electro-production process the spectra are peaked in the lower energy region; this is due to the fact that the $e^-$ production is a two steps process ($e^- \rightarrow \gamma \rightarrow e^+ + e^-$) so the primary energy is partially "wasted" in the photon production, therefore the positron photo-production can be seen as a more efficient process.

In fig. 11 the same plot as in fig. 10, but referring to the forward positron production in a cone of 10° aperture, are shown. The energy bin width is the same as in fig. 10.

The plots are less regular than the previous ones because of lower statistical accuracy.

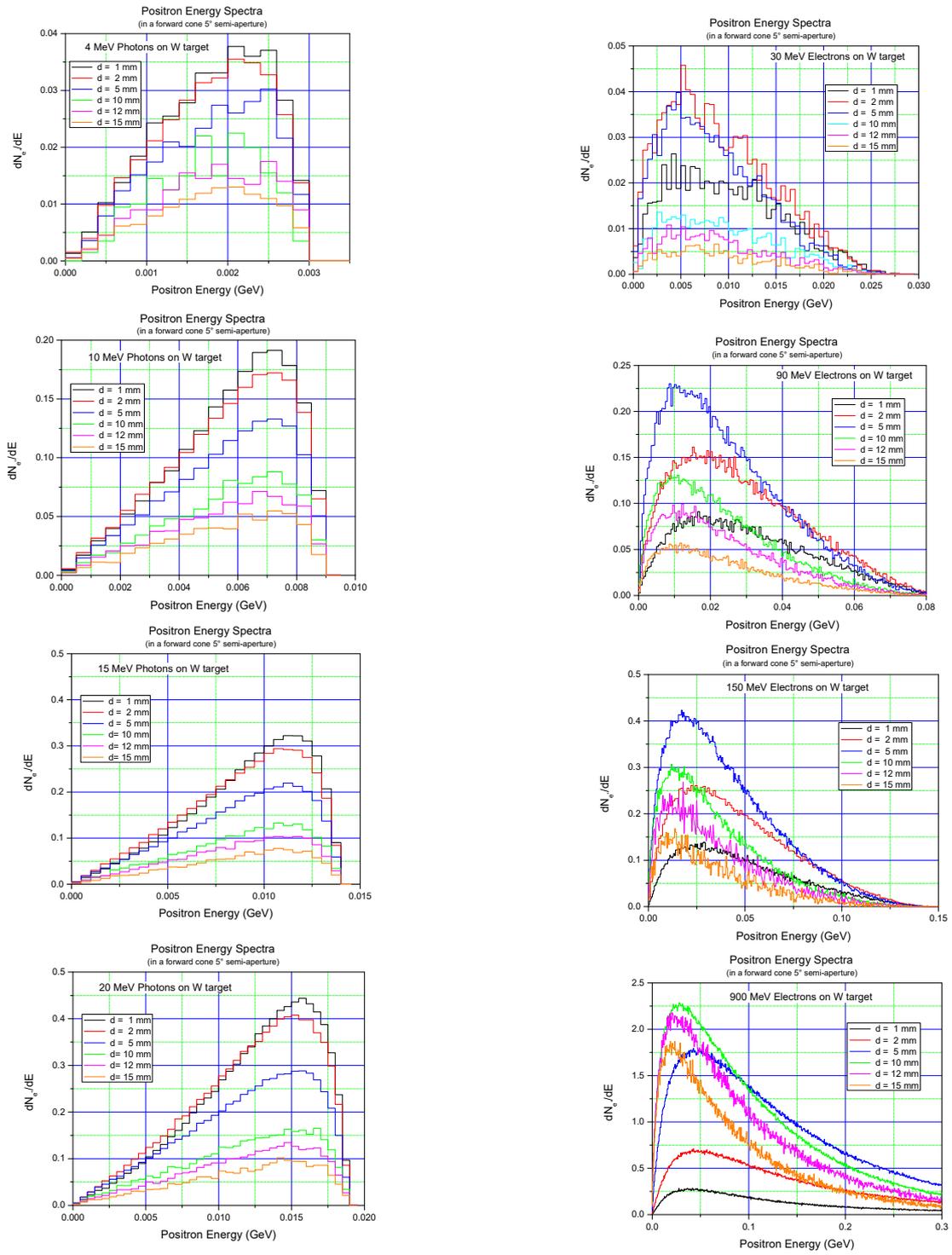

Figure 11: Positron spectrum in the forward direction (10° cone aperture)

## 4.4 Emittance and Brightness

After having determined the amount of the positron produced, let's evaluate the quality of the positron beam.

As a matter of fact in accelerator applications the positrons will be used as a "beam" and its intensity, energy, emittance and time structure must be known in order to match the beam to the experiment or to a beam transport system.

The rms emittance and normalized emittance of the produced positron beams is:

$$\varepsilon_{rms} = \sqrt{\left\langle x^2 \right\rangle \left\langle x'^2 \right\rangle - \left\langle xx' \right\rangle^2} \quad \text{and} \quad \varepsilon_n = \beta\gamma\varepsilon_{rms} \tag{4}$$

being $\beta$ and $\gamma$ the relativistic parameters.

Because of the wide energy spectrum, the emittance is evaluated by considering the positrons produced in energy slices of 0.5 MeV in all the cases.

Here we consider only the horizontal (x) emittance being the vertical one identical for symmetry reasons.

In fig. 12 the emittance and the normalized emittance of the emitted positrons as from 4 MeV primary photon beam are shown, the error associated with the emittance is inside the line thickness.

The top plot shows the energy distribution of the produced positrons, normalized to the total produced; this provides a statistical accuracy of the calculated emittance.

For example for positron energy $1 < E_{e+} < 1.25$ MeV the emittance and the normalized emittance, in meter·radians, for the 5 mm thick tungsten are $1.8 \times 10^{-4}$ and $5.5 \times 10^{-4}$ respectively, with these value calculated over about 36500 particles (the 12% of the total positrons produced).

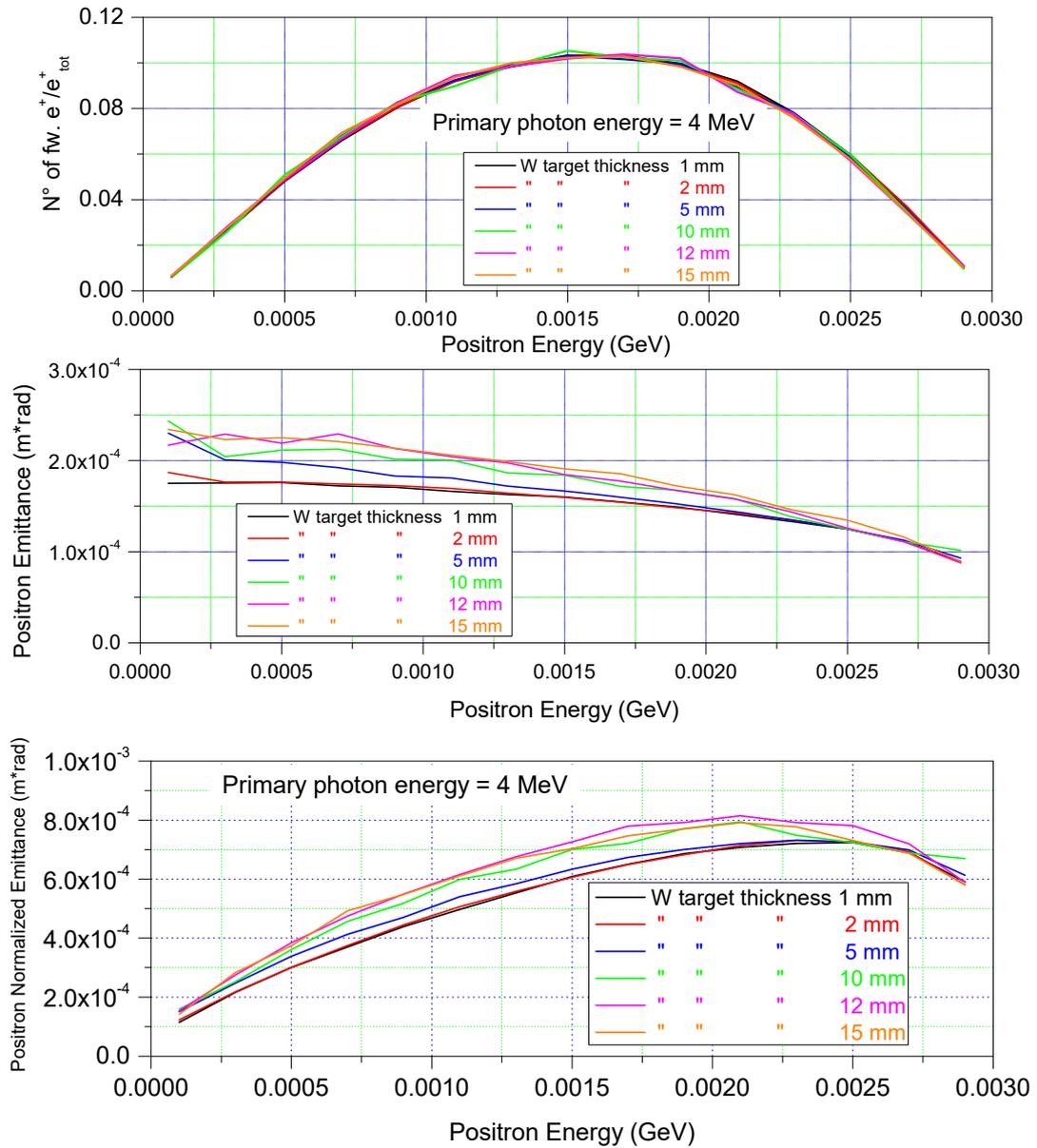

Figure 12: Positron emittance and normalized emittance vs. energy for positron produced by a 4 MeV primary photon beam on different thickness tungsten target. The top plot shows the particle/energy distribution, to indicate the statistical accuracy of the calculated emittances. (see text for details).

In fig. 13, 14 and 15 the same plots as in fig. 12 for 10 MeV,15 MeV and 20 MeV primary photons are shown respectively.

In fig. 15 the value of the emittance "a la Gruber" [20] is shown too. This is a proposed method to calculate the beam emittance of beam with large momentum spread. The case will be discussed in section 4.4.2.

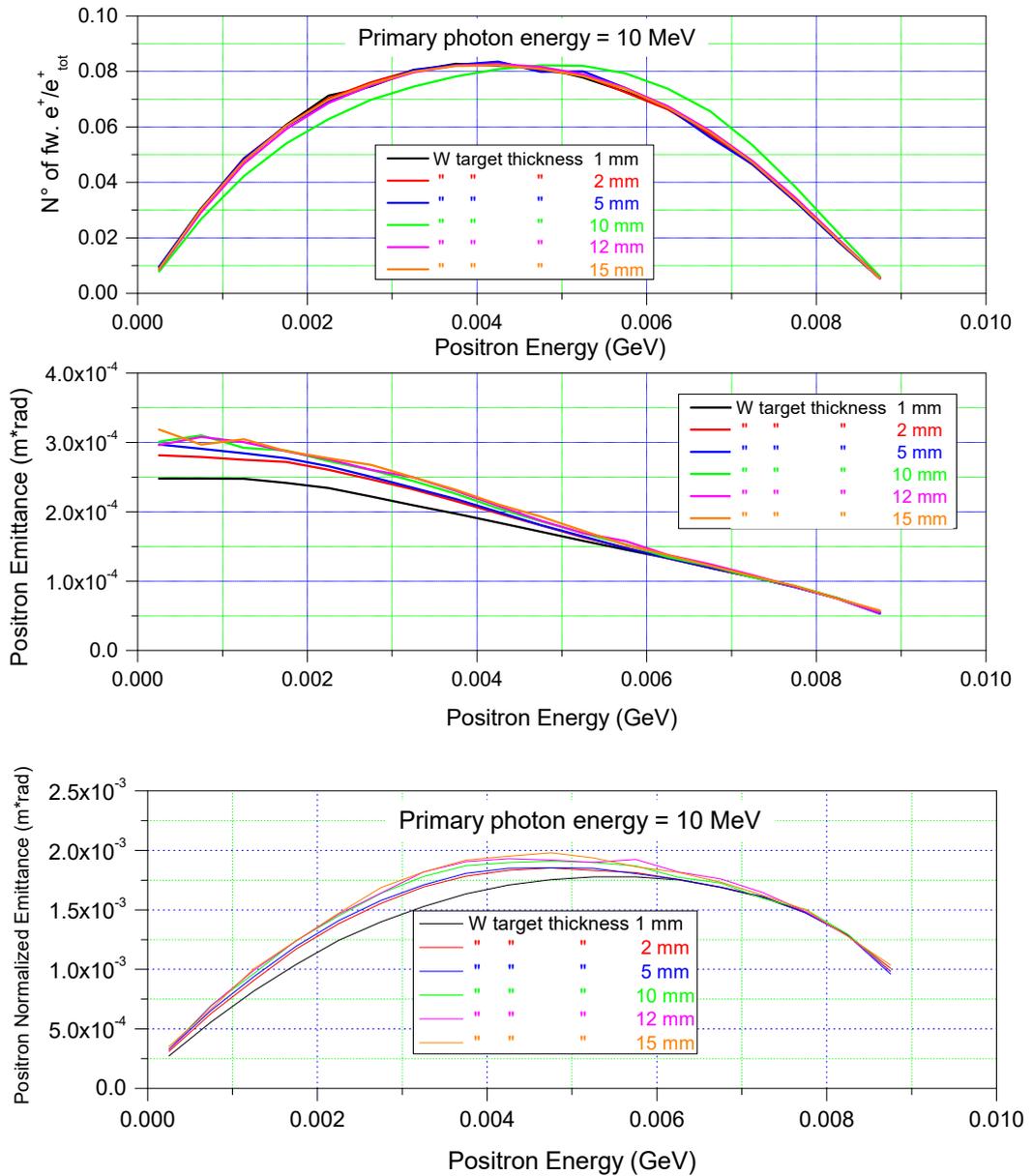

Figure 13: Positron emittance and normalized emittance vs. energy for positron produced by a 10 MeV primary photon beam on different thickness tungsten target. Top plot: particle/energy distribution. (see fig.12 and text for details).

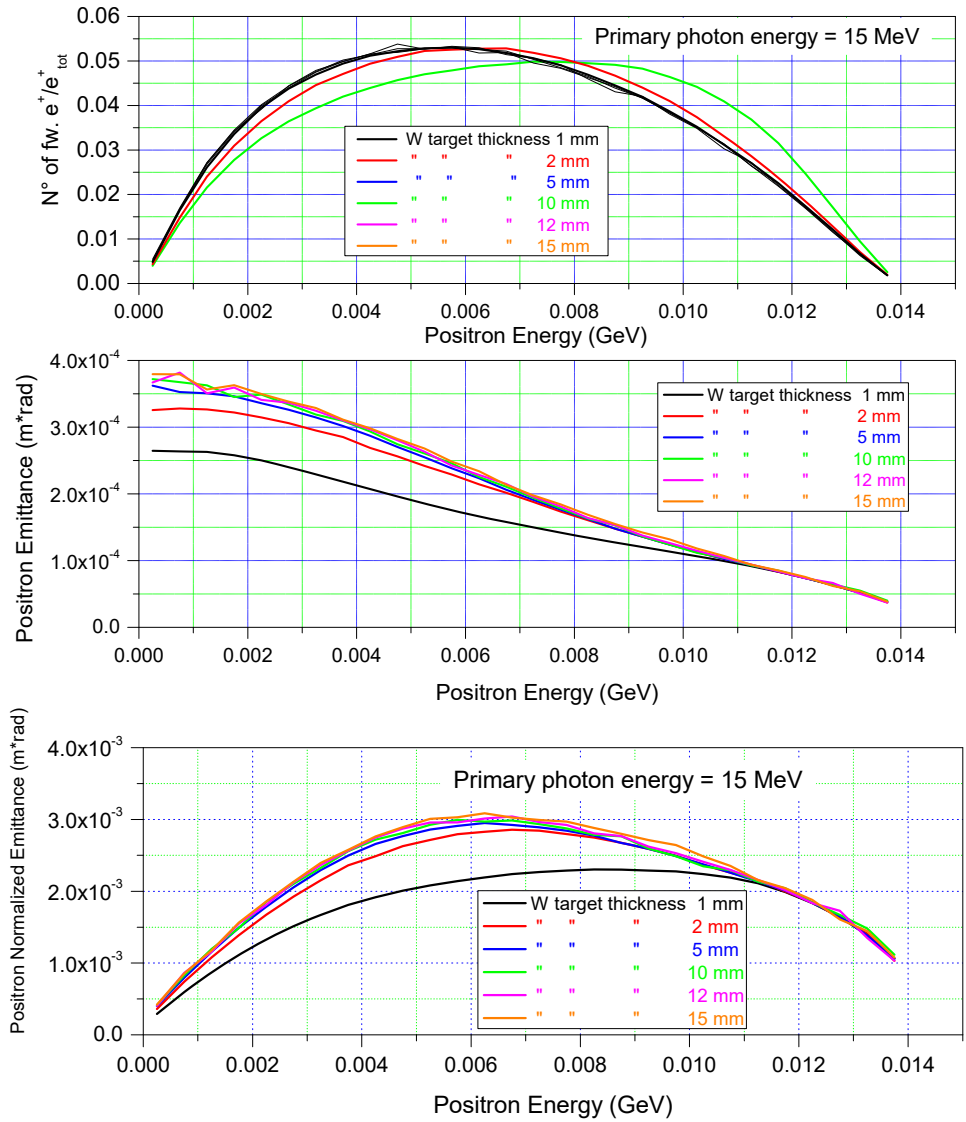

Figure 14: Positron emittance and normalized emittance vs. energy for positron produced by a 15 MeV primary photon beam on different thickness tungsten target. Top plot: particle/energy distribution. (see fig.12 and text for details).

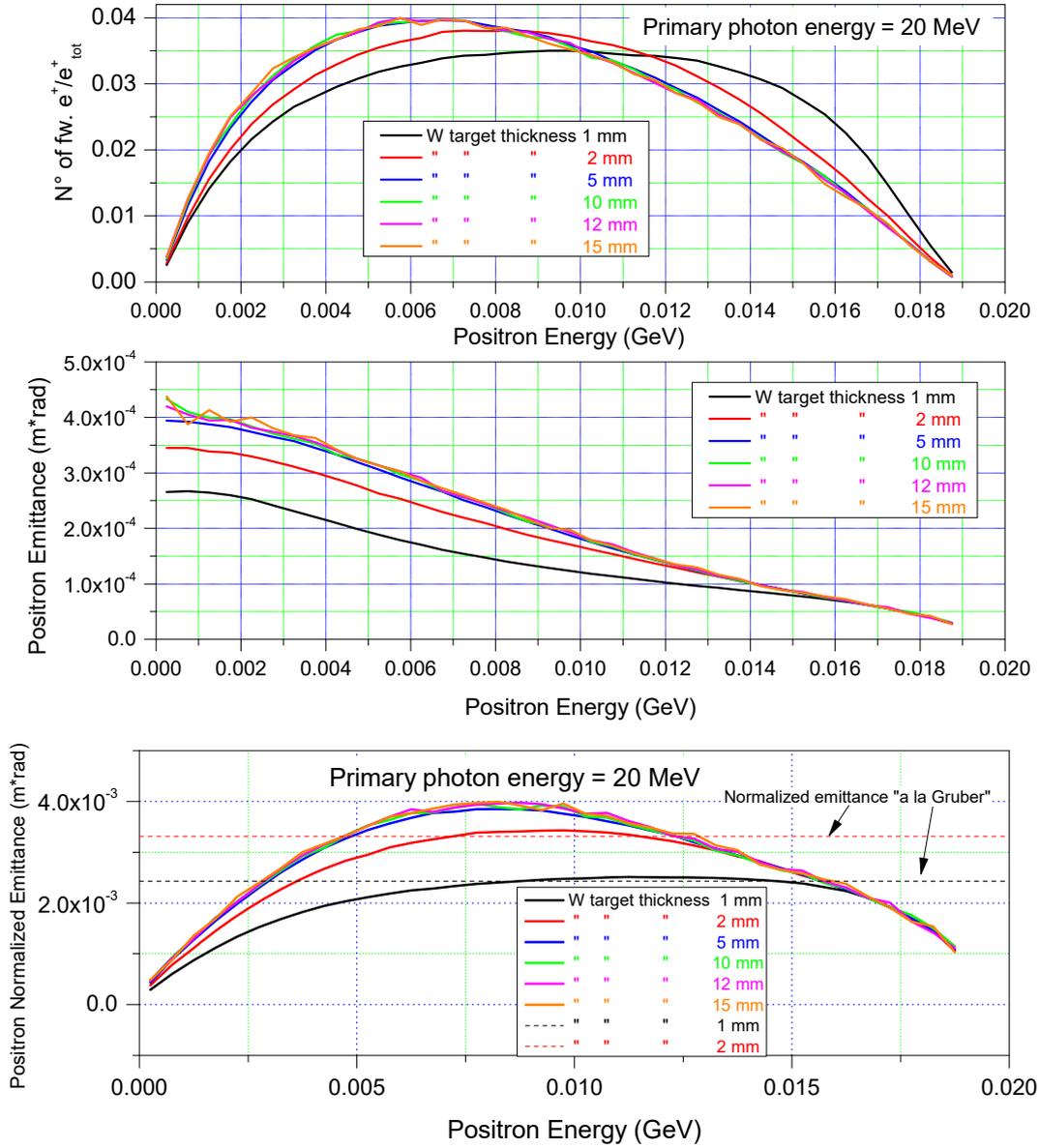

Figure 15: Positron emittance and normalized emittance vs. energy for positron produced by a 20 MeV primary photon beam on different thickness tungsten target. Top plot: particle/energy distribution. (see fig.12 and text for details). The normalized emittance "a la Gruber" for thickness of 1 mm (black dotted line) and 2 mm (red dotted line) is shown too (see section 4.4.2 for details).

In figs. 16-19 the emittance plots for the electron production with primary energy of 30, 90, 150 and 900 MeV are shown. Where the emittances are calculated over a low number of particles, high statistical fluctuations result.

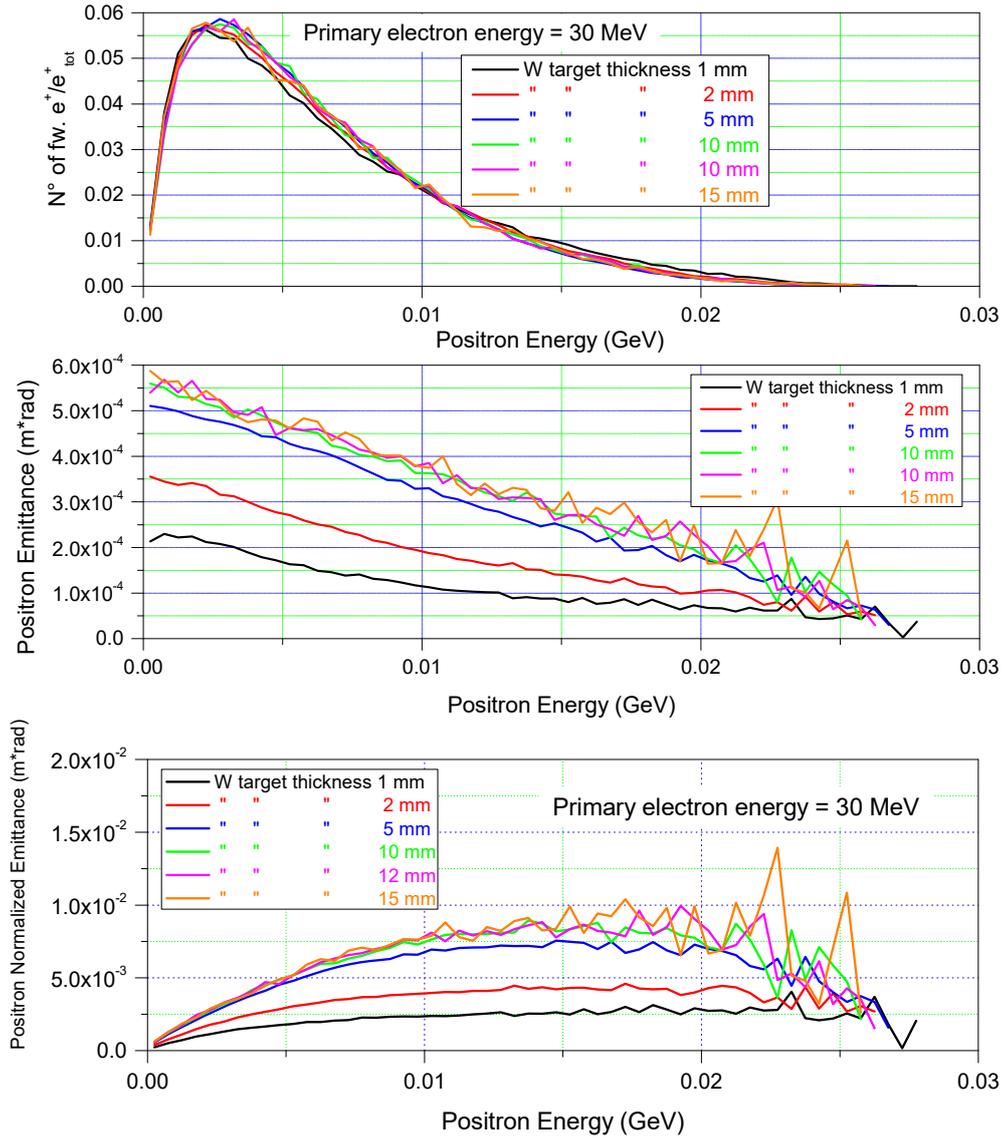

Figure 16: Positron emittance and normalized emittance vs. energy for positron produced by a 30 MeV primary electron beam on different thickness tungsten target. Top plot: particle/energy distribution. (see fig.12 and text for details).
The emittances are calculated over a low number of particles, giving high statistical fluctuations.

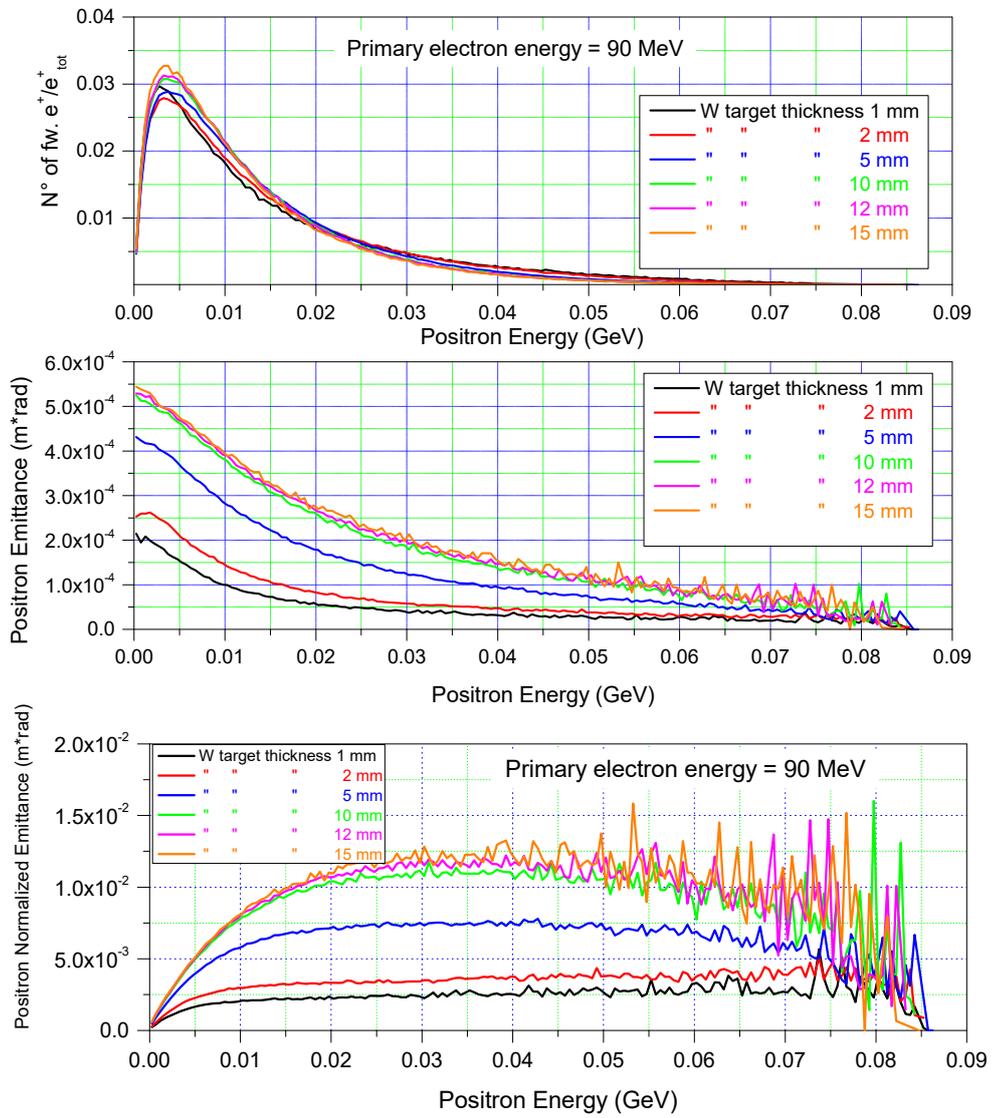

Figure 17: Positron emittance and normalized emittance vs. energy for positron produced by a 90 MeV primary electron beam on different thickness tungsten target. Top plot: particle/energy distribution. (see fig.12 and text for details).

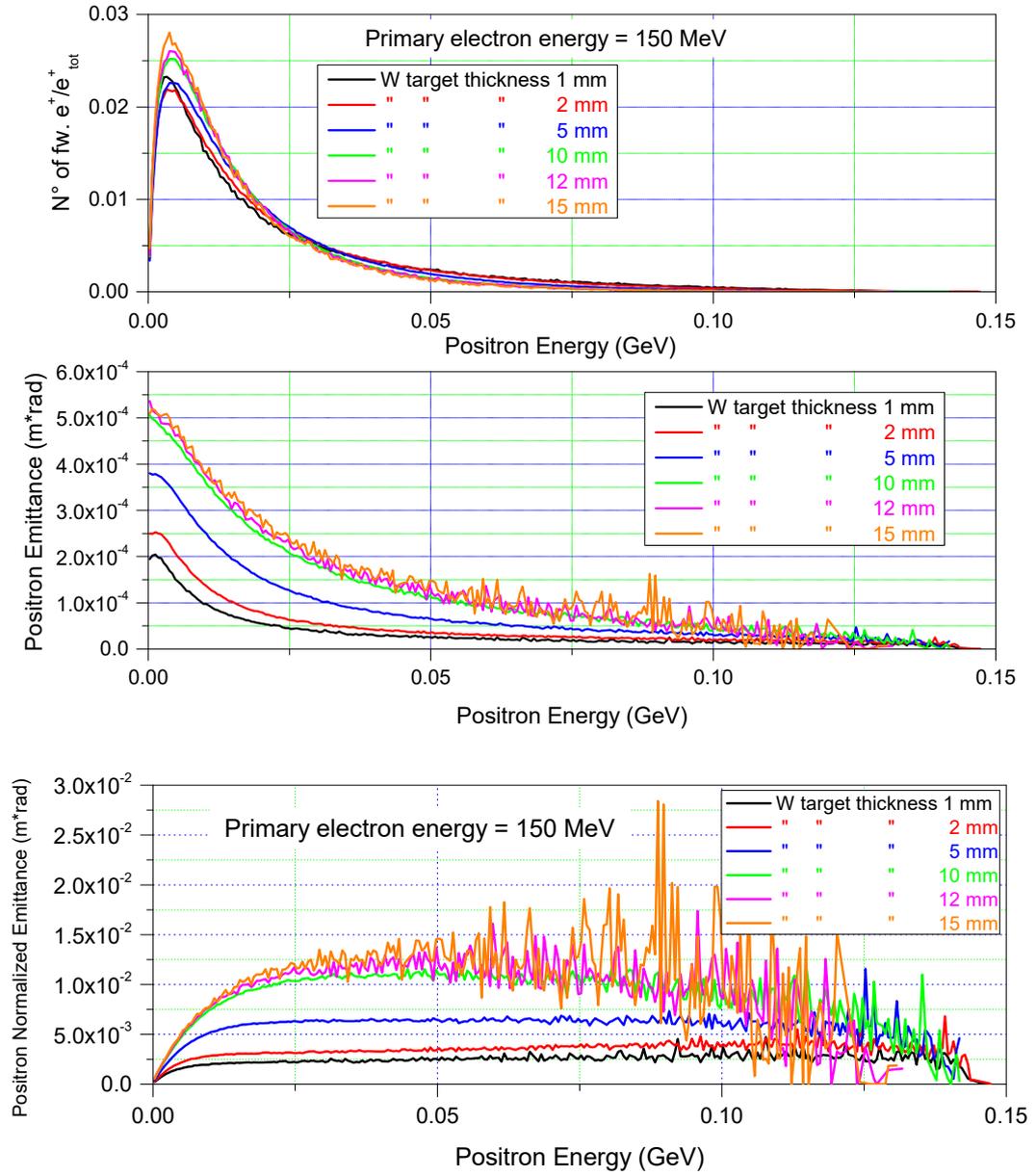

Figure 18: Positron emittance and normalized emittance vs. energy for positron produced by a 150 MeV primary electron beam on different thickness tungsten target. Top plot: particle/energy distribution. (see fig.12 and text for details).

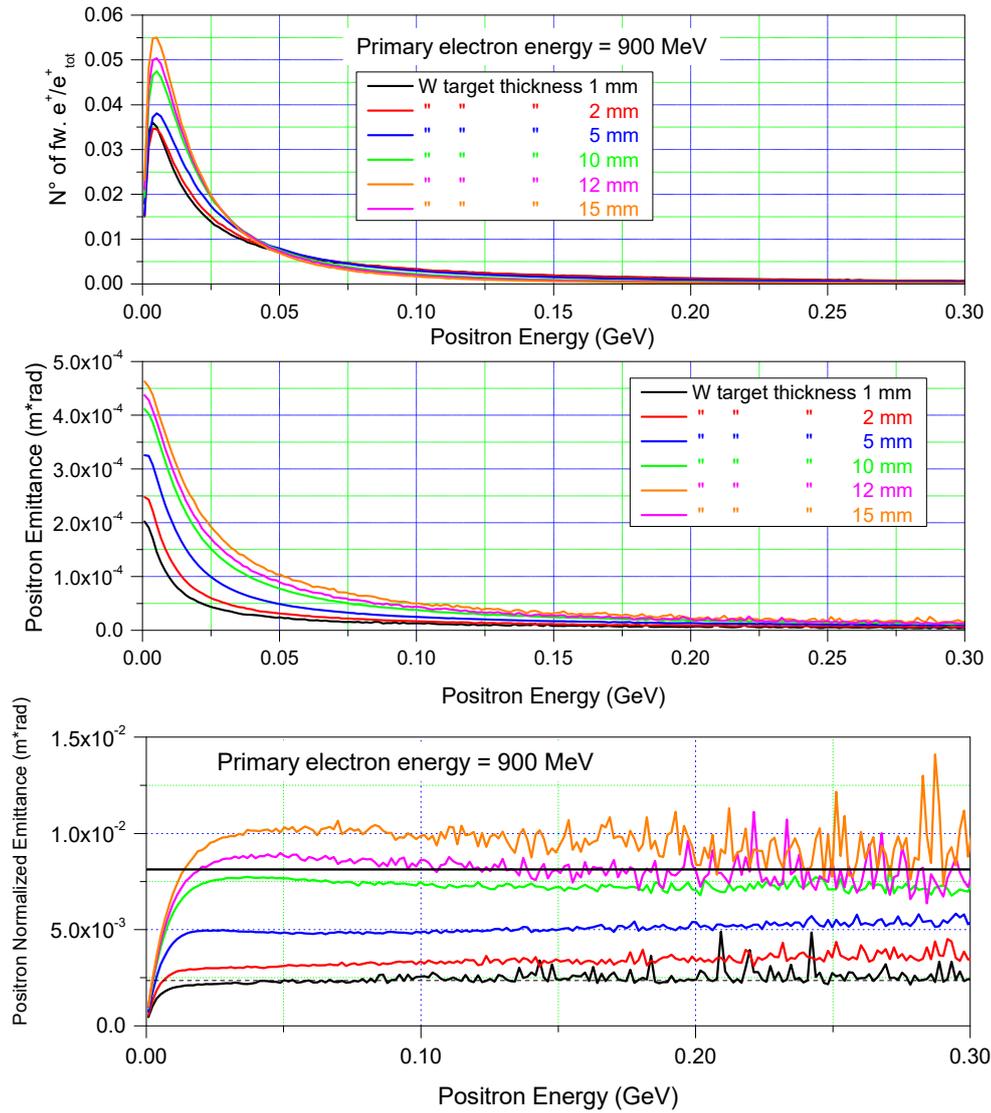

**Figure 19:** Positron emittance and normalized emittance vs. energy for positron produced by a 900 MeV primary electron beam on different thickness tungsten target. Top plot: particle/energy distribution. (see fig.12 and text for details).

The calculation of the normalized emittance "a la Gruber" for thickness of 1mm and 10 mm are $2.36 \times 10^{-3}$ m rad and $8.12 \times 10^{-3}$ m rad respectively, the first overlapping the horizontal part of the curves, the latter being just above the corresponding green line, so it is not reported on the plot.

The following tables show the values of the calculated emittance and normalized emittance for the energy interval where the maximum of the positron production occurs.

Tab. 2a refers to thickness 1 mm, 2 mm and 5 mm, while tab. 2b refers to thickness of 10 mm,12 mm and 15 mm. The error in the evaluation of the emittances is the percentage error.

Table.2a Emittance and normalized emittance for the energy interval where the maximum of the positron production occurs for tungsten thickness of 1 mm, 2 mm and 5 mm.

| 1.Target Thickness | | d = 1 mm | | | d = 2 mm | | | d = 5 mm | | |
|---|---|---|---|---|---|---|---|---|---|---|
| Primary Beam | Primary Energy (MeV) | $e^+$ Energy (MeV) | $\varepsilon$ ± err% (m rad) | $\varepsilon_n$ ± err% (m rad) | $e^+$ Energy (MeV) | $\varepsilon$ ± err% (m rad) | $\varepsilon_n$ ± err% (m rad) | $e^+$ Energy (MeV) | $\varepsilon$ ± err% (m rad) | $\varepsilon_n$ ± err% (m rad) |
| Photons | 4 | 1.625±0.25 | 1.56E-04 ± 8E-04 | 6.33E-04 ± 8E-04 | 1.625±0.25 | 1.56E-04 ± 2E-03 | 6.34E-04 ± 2E-03 | 1.625±0.25 | 1.62E-04 ± 6E-03 | 6.59E-04 ± 6E-03 |
| | 10 | 4.75±0.25 | 1.71E-04 ± 2E-04 | 1.75E-03 ± 2E-04 | 4.25±0.25 | 1.98E-04 ± 3E-04 | 1.83E-03 ± 3E-04 | 4.25±0.25 | 2.0E-04 ± 3.E-04 | 1.85E-03 ± 3E-04 |
| | 15 | 7.25±0.25 | 1.50E-04 ± 8E-04 | 2.27E-03 ± 8E-04 | 6.75±0.25 | 2.02E-04 ± 1E-04 | 2.86E-03 ± 1E-04 | 6.25±0.25 | 2.24E-04 ± 3.E-05 | 2.95E-03 ± 3E-05 |
| | 20 | 9.25±0.25 | 1.29E-04 ± 7E-04 | 2.46E-03 ± 7E-04 | 7.25±0.25 | 2.19E-04 ± 3E-04 | 3.32E-03 ± 3E-04 | 6.75±0.25 | 2.66E-04 ± 1E-04 | 3.77E-03 ± 1E-04 |
| Electrons | 30 | 2.25±0.25 | 2.12E-04 ± 6E-02 | 1.12E-03 ± 6E-02 | 2.25±0.25 | 3.35E-04 ± 8E-03 | 1.78E-03 ± 8E-03 | 2.75±0.25 | 4.76E-04 ± 5E-03 | 3.0E-03 ± 5E-03 |
| | 90 | 2.75±0.25 | 1.86E-04 ± 2E-02 | 1.17E-03 ± 2E-02 | 3.25±0.25 | 2.43E-04 ± 9E-03 | 1.77E-03 ± 9E-03 | 3.75±0.25 | 3.92E-04 ± 3E-03 | 3.24E-03 ± 3E-03 |
| | 150 | 3.25±0.25 | 1.79E-04 ± 1E-02 | 1.31E-03 ± 1E-02 | 3.25±0.25 | 2.35E-04 ± 1E-02 | 1.72E-03 ± 1E-02 | 3.75±0.25 | 3.53E-04 ± 3E-03 | 2.93E-03 ± 3E-03 |
| | 900 | 4.25±0.25 | 1.65E-04 ± 9E-03 | 1.53E-03 ± 9E-03 | 4.25±0.25 | 2.14E-04 ± 5E-03 | 1.98E-03 ± 5E-03 | 5.25±0.25 | 2.84E-04 ± 3E-04 | 3.18E-03 ± 3E-04 |

Table.2b Emittance and normalized emittance for the energy interval where the maximum of the positron production occurs for tungsten thickness of 10 mm, 12 mm and 15 mm.

| Target Thickness | | d = 10 mm | | | d = 12 mm | | | d = 15 mm | | |
|---|---|---|---|---|---|---|---|---|---|---|
| Primary Beam | Primary Energy (MeV) | e+ Energy (MeV) | ε ± err% (m rad) | εn ± err% (m rad) | e+ Energy (MeV) | ε ± err% (m rad) | εn ± err% (m rad) | e+ Energy (MeV) | ε ± err% (m rad) | εn ± err% (m rad) |
| Photons | 4 | 1.625±0.25 | 1.75E-04 ± 1E-02 | 7.11E-04 ± 1E-02 | 1.875±0.25 | 1.70E-04 ± 3E-02 | 7.74E-04 ± 3E-02 | 1.625±0.25 | 1.88E-04 ± 5E-03 | 7.63E-04 ± 5E-03 |
| | 10 | 3.75±0.25 | 2.26E-04 ± 5E-03 | 1.87E-03 ± 5E-03 | 4.25±0.25 | 2.08E-04 ± 1E-02 | 1.93E-03 ± 1E-02 | 4.25±0.25 | 2.11E-04 ± 8E-03 | 1.95E-03 ± 8E-03 |
| | 15 | 5.75±0.25 | 2.45E-04 ± 9E-03 | 3.00E-03 ± 9E-03 | 4.75±0.25 | 2.8E-04 ± 3E-02 | 2.87E-03 ± 3E-02 | 5.75±0.25 | 2.48E-04 ± 1E-02 | 3.03E-03 ± 1E-02 |
| | 20 | 6.75±0.25 | 2.67E-04 ± 8E-03 | 3.78E-03 ± 8E-03 | 5.75±0.25 | 3.03E-04 ± 2E-02 | 3.7E-03 ± 2E-02 | 5.75±0.25 | 3.03E-04 ± 2E-03 | 3.7E-03 ± 2E-03 |
| Electrons | 30 | 2.75±0.25 | 5.07E-04 ± 5E-02 | 3.20E-03 ± 5E-02 | 3.25±0.25 | 4.97E-04 ± 6E-02 | 3.62E-03 ± 6E-02 | 3.75±0.25 | 4.75E-04 ± 5E-02 | 3.93E-03 ± 5E-02 |
| | 90 | 3.25±0.25 | 4.86E-04 ± 3E-03 | 3.55E-03 ± 3E-03 | 3.25±0.25 | 4.97E-04 ± 3E-03 | 3.62E-03 ± 3E-03 | 3.75±0.25 | 4.95E-04 ± 1E-02 | 4.1E-03 ± 1E-02 |
| | 150 | 3.75±0.25 | 4.65E-04 ± 7E-04 | 3.85E-03 ± 7E-04 | 3.75±0.25 | 4.86E-04 ± 4E-02 | 4.03E-03 ± 4E-02 | 3.75±0.25 | 5.0E-04 ± 1E-02 | 4.14E-03 ± 1E-02 |
| | 900 | 5.25±0.25 | 3.63E-04 ± 6E-04 | 4.07E-03 ± 6E-04 | 5.25±0.25 | 3.87E-04 ± 3E-03 | 4.35E-03 ± 3E-03 | 4.25±0.25 | 4.34E-04 ± 1E-03 | 4.02E-03 ± 1E-03 |

The plots of the data reported in the previous tables are shown in fig.20.

In fig. 20a the emittances vs. the primary energy (both photon and electrons) calculated at the positron energy where the maximum of production occurs (+/- 250 keV), as reported in the tables are plotted; fig. 20b shows the same emittance values vs. the tungsten target thickness.

As we can see the emittance is generally lower for photo-production, this because of the two stages process of the electro-production, generating a larger shower of the bremsstrahlung photons respect to the primary photons (see fig. 6).

By looking at the emittance by photo-production we can see that the emittance increases with the primary energy with exception of the 1 mm thickness, in this case the decreasing of the emittance with the energy is probably due to the increasing of the emitted positron energy and the consequent increase in the longitudinal momentum (larger for the case of 1 mm respect to the one for the other thickness, see tab. 2a) and the lower scattering effect due to the small thickness of the target.

In case of electro-production the emittance vs the primary energy decreases, because of the increasing of the longitudinal momentum of the produced positrons.

The increasing of the emittance with the target thickness, both for photo-production and electro-production, is due to the scattering of the primaries while travelling inside the target and the consequent generation of the secondary positrons in a wider surface, and by the effect of multiple scattering of the secondary positrons.

The decreasing of the emittance with the thickness in case of electro-production for the 30 MeV electrons primary energy (green trace and rhombic symbol in fig. 20b) may be due to the increasing of the energy where the emittance is calculated (see tab. 2b).

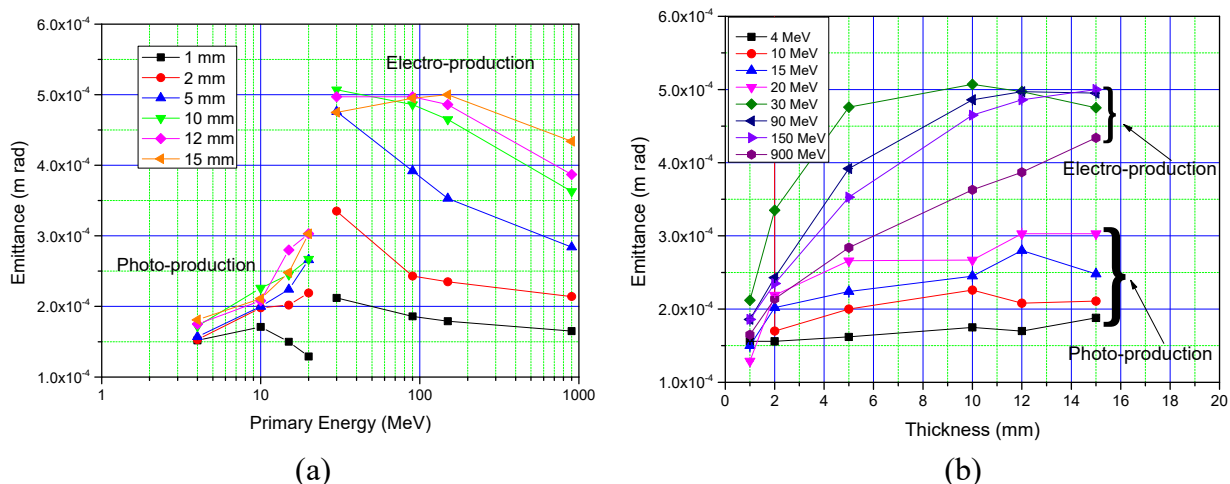

(a)                                   (b)

Figure 20: Positron emittance at the maximum yield (+/- 0.25 MeV) for photo-production and electro-production vs. primary energy (a). Positron emittance vs. target thickness (b)

Anyway with the increasing of the emittance it must be taken into account the increase of the total number of the positrons produced (it increases with the energy, see figs.5), so a trade-off between intensity vs. quality of the secondary positron "beam" produced must be considered, as shown in tab. 3.

In tab. 3a and tab. 3b the values of the total emitted positrons (per primary), and the corresponding emittance at the energy of the maximum yield together with the number of positrons (respect to the total produced, and respect to the primary) are shown for the case of best yield and the case of best emittance, for photo-production (tab.3a) and electro-production (tab. 3b.).

Table 3a. Best yield vs. best emittance for photo-production. $(dN_{e+}(E)/N_{e+})$ is the fraction of positron in the energy range respect to the total produced. $(dN_{e+}(E)/N_{pr})$ is the same number respect to the primary.

| Photo-production | | | | | | | |
|---|---|---|---|---|---|---|---|
| | Primary Energy (MeV) | Thickness (mm) | Total $e^+$/prim | Emittance (m rad) | $e^+$ Energy (MeV) | $dN_{e+}(E)/N_{e+}$ | $dN_{e+}(E)/N_{pr}$ |
| Best Yield | 20 | 2 | 9.24E-02 | 2.19E-04 | 7.25±0.25 | 3.80295E-02 | 3.51398E-03 |
| Best Emittance | 20 | 1 | 6.91E-02 | 1.29E-04 | 9.25±0.25 | 3.50205E-02 | 2.42089E-03 |

Table 3b. Best yield vs. best emittance for electro-production

| Electro-production | | | | | | | |
|---|---|---|---|---|---|---|---|
| | Primary Energy (MeV) | Thickness (mm) | Total $e^+$/prim | Emittance (m rad) | $e^+$ Energy (MeV) | $dN_{e+}(E)/N_{e+}$ | $dN_{e+}(E)/N_{pr}$ |
| Best Yield | 900 | 10 | 2.27 | 3.63E-04 | 5.25E-03 | 1.59180E-02 | 3.64896E-02 |
| Best Emittance | 900 | 1 | 9.03E-02 | 1.65E-04 | 4.25E-03 | 1.20110E-02 | 1.17870E-03 |

From the data of tab. 3a it can be seen that for photo-production the situation of best production yield respect to the best emittance value has a ratio of about 1.34 in the yield with a reduction of a factor about 0.6 in the emittance.

In the case of electro-production from tab. 3b the best production yield vs. the yield for the best emittance is about a factor of 25 with a reduction in the emittance of about 2.2.

The best choice depends on the specific application.

Let's now analyze the emittance of the produced positrons in the cases where the maximum of the photo-production and electro-production occurs (as reported in tab.1) i.e. 20 MeV primary photons on 2 mm Tungsten, and 900 MeV electrons on 10 mm Tungsten.

In fig. 21a and 21b the transverse phase space in these two cases is reported. The plots refer to the energy where the maximum positron yield occurs, i.e. 7.25 +/- 0.25 MeV for the photo-production and 5.25 +/- 0.25 MeV for the electron-production. As explained before in the case of electro-production the beam is wider because of the two stages process, leading to a more spread shower.

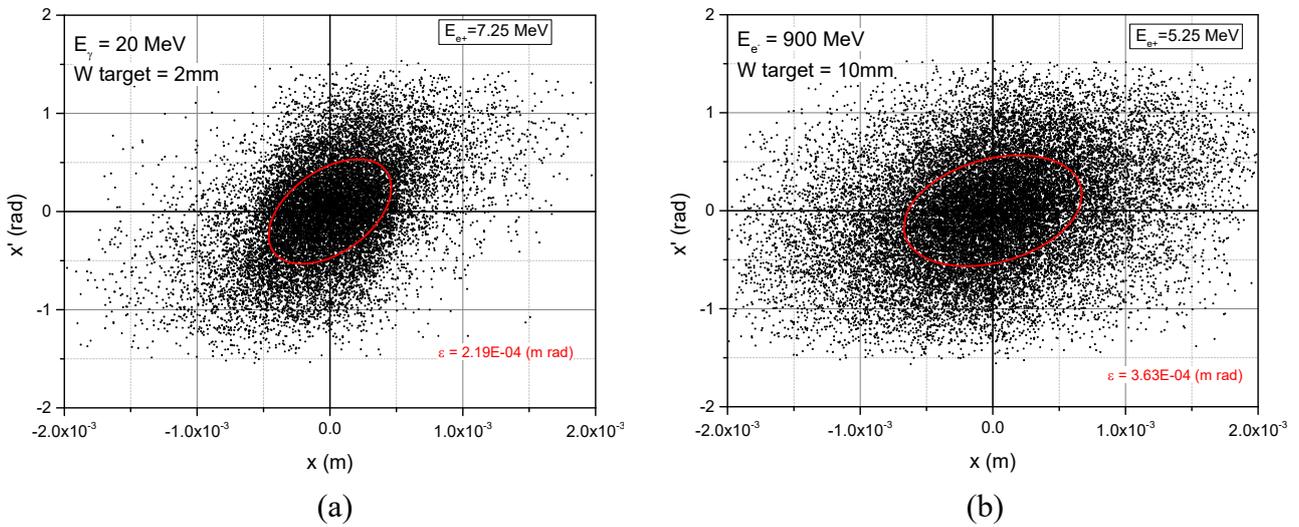

(a)                    (b)

Figure 21: (a) Positron phase space at the exit of the 2 mm tungsten target from 20 MeV primary photons at 7.25+/- 0.25 MeV (maximum yield), (b) the same plot for 900 MeV electro-production on 10 mm target thickness for 5.25 +/- 0.25 MeV positrons.

In fig. 22a the plot of the emittances of the positrons produced by 20 MeV photon beam on a 2 mm W target is shown for three different energies: 7.25 MeV (7<E<7.5) that is the energy at which the maximum of the secondary positrons are produced (as told before) and a lower and higher energy (0.75 MeV and 18.5 MeV respectively, all +/- 0.25 MeV).

Fig. 22b is the same as fig 20a but in case of electro-production by a 900 MeV electron beam on a 10 mm W target at 1.25 MeV, 5.25 MeV and 49.75 MeV (all +/- 0.25 MeV).

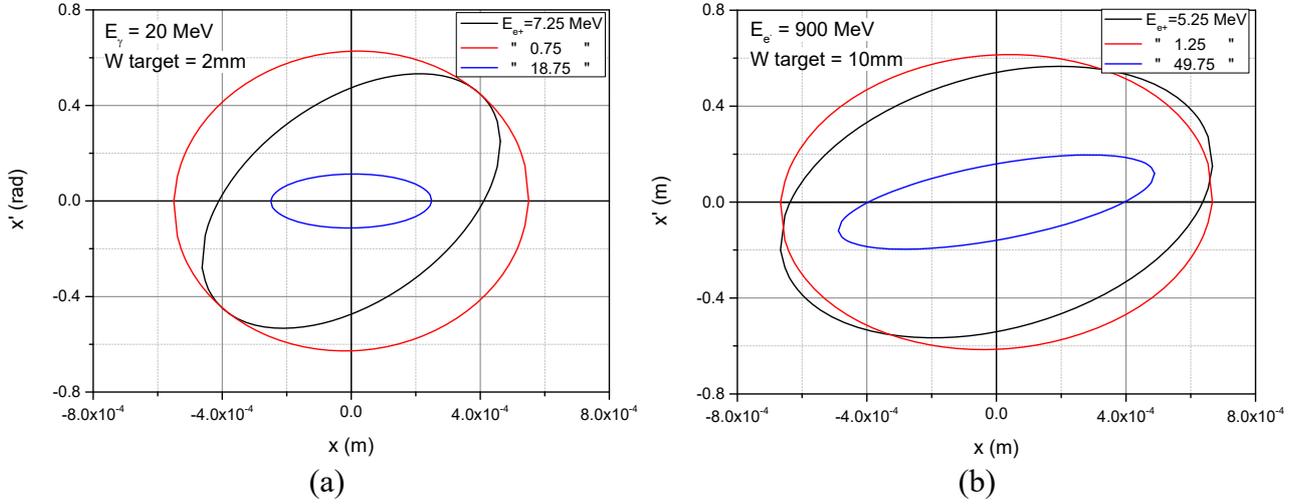

Figure 22: (a) Positron phase space at the exit of the 2 mm tungsten target from 20 MeV primary photons for three different energies, (b) the same plot for 900 MeV electro-production on 10 mm target thickness.

From the plots of fig. 22 is evident the decreasing of the emittance with as the energy of the positrons.

Tab. 4 reports the data corresponding to the emittance plots of fig. 22.

In the last columns the number of positrons in the considered energy range over the number of the total positrons and the same number of positrons over the number of primary is reported.

It can be seen that in both photo-production and electro-production the emittance decrease with the positron energy, it follows, as analyzed before, that the optimum for a positron source is a trade-off between the high production yield and the good quality of the beam (low emittance).

Table 4

| Primary Beam | Primary Energy /W thickness | Positron Energy (MeV) | ε (m rad) ± err% | εₙ (m rad) ± err% | dN_{e+}(E)/N_{e+} | dN_{e+}(E)/N_{pr} |
|---|---|---|---|---|---|---|
| Photons | 20MeV /2mm | 0.75 | 3.45E-04 ± 0.3 | 7.78E-04 ± 0.4 | 9.75949E-03 | 9.01790E-04 |
| | | 7.25 | 2.19E-04 ± 0.3 | 3.32E-03 ± 0.3 | 3.80295E-02 | 3.51398E-03 |
| | | 18.75 | 2.82E-05 ± 1.3 | 1.06E-03 ± 1.3 | 9.91219E-04 | 9.15900E-05 |
| Electrons | 900MeV /10mm | 1.25 | 4.09E-04 ± 0.1 | 1.35E-03 ± 0.1 | 1.02947E-02 | 2.35996E-02 |
| | | 5.25 | 3.63E-04 ± 0.1 | 4.08E-03 ± 0.1 | 1.59017E-02 | 3.64530E-02 |
| | | 49.75 | 7.81E-05 ± 0.5 | 7.68E-03 ± 0.5 | 2.53145E-03 | 5.80310E-03 |

*4.4.1 Effect of the primary beam spot dimension on the emittance*

As from fig. 9 the fraction of the positrons emitted in a forward narrow cone ± 5° wide spans from 2% to 6% for the photo-production and from 2% to 45% for the electron production; in a cone ± 0.32° wide we have 0.01% to 0.03% for the photo-production and 0.01% to 1.6% for the electron production.

The positrons emitted from the target have a wide angular spectrum, with divergence of about 0.6 rad at the lowest energy and decreasing as the energy increase, down to above 0.1-0.3 rad for the photon production and 0.2–0.02 for the electron production at the maximum energy, depending on the target thickness (see figs. 21 and 22); these values of divergence are independent on the dimension of the primary beam.

Conversely the beam transverse dimension is strongly related to the dimension of the primary beam and a narrow primary beam lead to a narrower positron emission and, consequently, a reduced transverse emittance.

In tab 5 the values of the emittance and normalized emittance for photo-production and electro-production in case of thin and thick tungsten target (1mm and 15 mm) for different primary beam radii are reported.

The ratio of the produced positrons respect to the total positron produced and the positrons respect to the primary intensity is reported too.

Table 5 Emittance and normalized emittance for the thinnest (1mm) and Thickest (15 mm) tungsten target
for different primary beam radii

| W Target Thickness = 1 mm | | | | | | | | | |
|---|---|---|---|---|---|---|---|---|---|
| | | Photons (E$_\gamma$ = 20 MeV) | | | | Electrons (E$_e$ = 900 MeV) | | | |
| Primary Beam radius (cm) | e$^+$ Energy ± 0.25 (MeV) | ε ± err% (m rad) | ε$_n$ ± err% (m rad) | dN$_{e+}$(E)/N$_{e+}$ | dN$_{e+}$(E)/N$_{pr}$ | e$^+$ Energy (MeV) | ε ± err% (m rad) | ε$_n$ ± err% (m rad) | dN$_{e+}$(E)/N$_{e+}$ | dN$_{e+}$(E)/N$_{pr}$ |
| 0.005 | 9.75 | 6.00E-05 ± 1.4 | 1.20E-03 ± 1.4 | 3.51514E-02 | 2.42910E-03 | 3.75 | 1.04E-04 ± 5.0 | 8.62E-04 ± 5.0 | 1.24767E-02 | 1.22600E-03 |
| 0.05 | 9.75 | 1.20E-04 ± 0.7 | 2.48E-03 ± 0.7 | 3.51093E-02 | 2.42730E-03 | 4.25 | 1.65E-04 ± 0.9 | 1.52E-03 ± 0.9 | 1.20110E-02 | 1.17870E-03 |
| 0.5 | 8.75 | 1.15E-03 ± 0.6 | 2.21E-02 ± 0.6 | 3.52748E-02 | 2.44340E-03 | 3.75 | 1.40E-03 ± 2.0 | 1.10E-02 ± 2.0 | 1.23342E-02 | 1.20600E-03 |
| W Target Thickness = 15 mm | | | | | | | | | | |
| 0.005 | 6.25 | 2.56E-04 ± 3.5 | 3.38E-03 ± 3.5 | 4.23620E-02 | 1.05600E-03 | 5.25 | 4.07E-04 ± 1.1 | 4.57E-03 ± 1.1 | 5.48507E-02 | 1.06650E-01 |
| 0.05 | 6.75 | 2.72E-04 ± 1.1 | 3.85E-03 ± 1.1 | 3.97738E-02 | 9.53100E-04 | 5.25 | 4.15E-04 ± 0.3 | 4.66E-03 ± 0.3 | 5.50088E-02 | 1.05727E-01 |
| 0.5 | 4.25 | 1.71E-03 ± 4.6 | 1.59E-02 ± 4.6 | 4.05440E-02 | 1.10900E-03 | 3.75 | 1.86E-03 ± 0.8 | 1.54E-02 ± 0.8 | 6.56482E-02 | 1.57660E-01 |

In fig. 23 the transverse emittance, the normalized emittance and the r.m.s. radius of the emitted positrons are plotted for the case of 20 MeV primary photon beam hitting a 1mm tungsten target (left) and 15 mm tungsten target (right) with 0.05 mm, 0.5 mm and 5 mm primary beam radius.

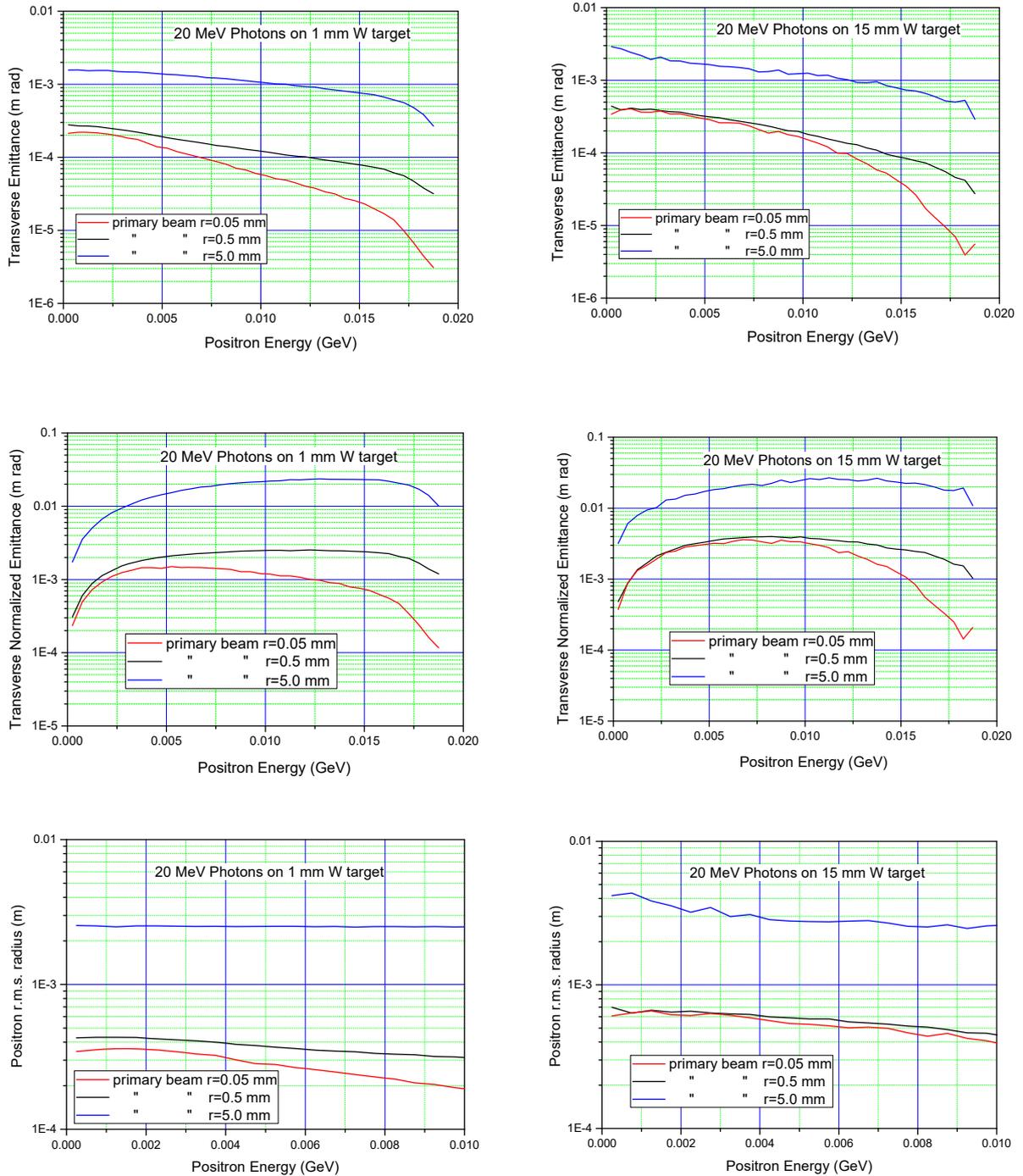

Figure 23: Positron emittance (top) , normalized emittance (middle) and beam radius (bottom) vs. energy of the for positron produced by a 20 MeV primary photon beam on 1 mm tungsten target (left plots) and 15 mm target (right plots).

In fig. 24 the same plots of fig. 23 for the case of 900 MeV primary electron beam are shown.

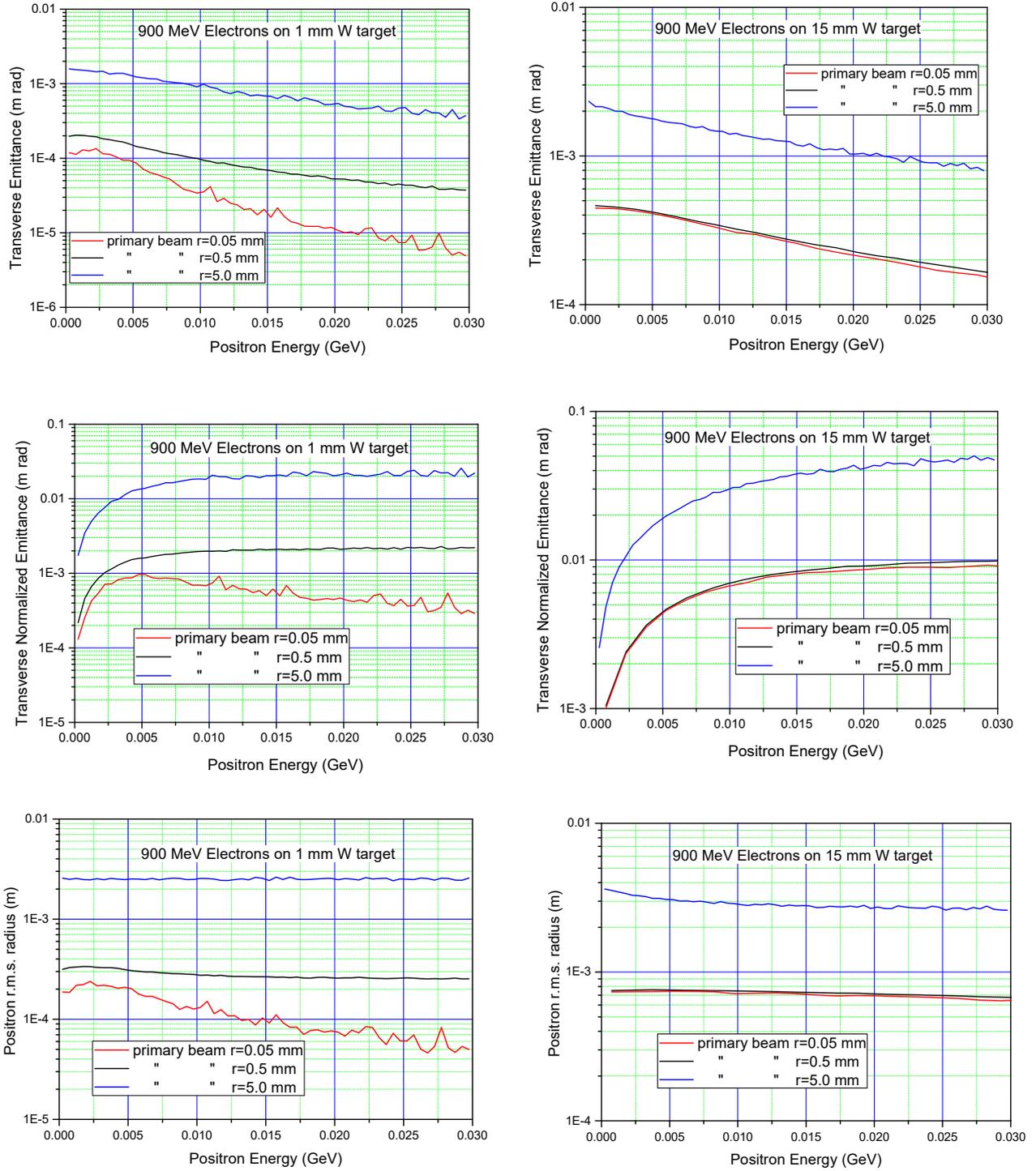

Figure 24: (a)Positron emittance (top), normalized emittance (middle) and beam radius (bottom) vs. energy of the for positron produced by a 900 MeV primary electron beam on 1 mm tungsten target (left plots) and 15 mm target (right plots).

The same considerations on the statistic accuracy on the plots done for the previous plots are still valid.

From the plots we can see that the positrons emitted at the higher energies by photo-production reflects the primary beam dimension (as a matter of fact their emittance and normalized emittance of the different primary radii, at high energy, differ of about a factor 10), while at low energy the positron beam has a larger radius, this probably because of the scattering occurring inside the target before being emitted. As a matter of fact the "age" of the positrons (i.e. the time between their generation and their exit from the target) is higher at lower energy, as shown in fig 25.

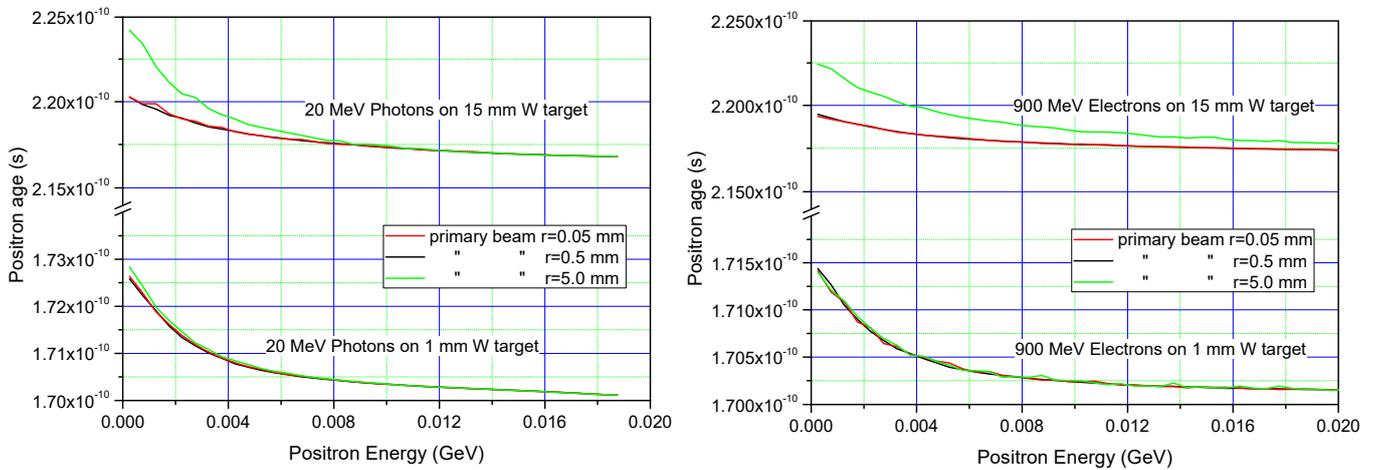

Figure 25: Positron age, i.e. the time lived inside the target for photo-production (left) and electro-production (right).

From fig. 25 the age of the positron by electro-production is slightly lower than the photo-production and this difference is higher at lower energies. This difference is due to the location inside the target where the positrons are produced, closer to the target entry for the photo-production case while the production is closer to the target exit for the electro-production, as shown in figs. 6 and 7. Both for photo-production and electro-production the different age of the positrons is evident only for the largest primary beam for thick target.

In fig.26 the "age" of the positrons is plotted versus the energy for the different target thickness; the continuous lines refer to the photo-production, while the dotted lines are for the electron-production. It is not evident the difference between photo vs. electro-production; more evident is the difference among the different thickness.

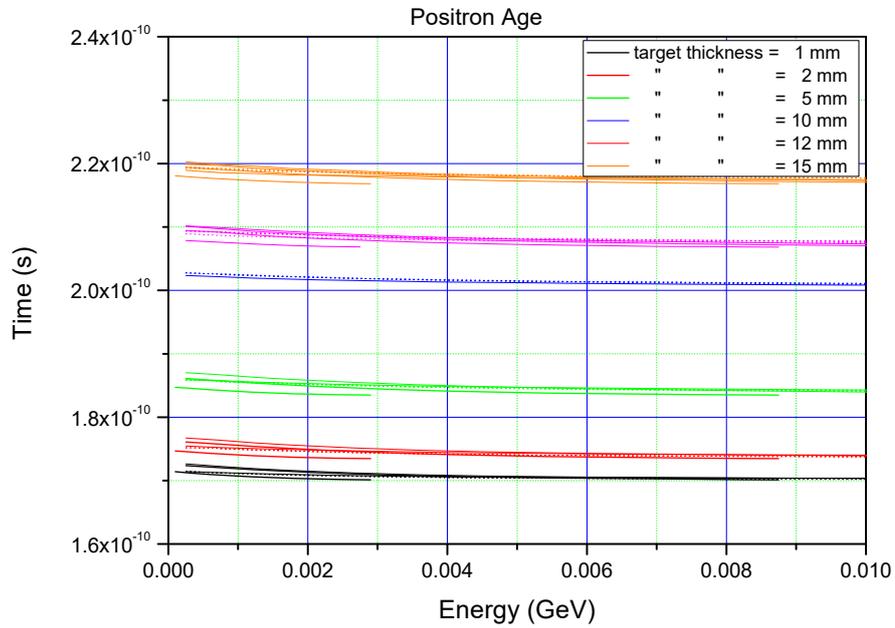

Figure 26: Positron age, i.e. the time lived inside the target for all the studied cases. Different colours for different thickness, continuous line for photo-production, dotted lines for electro-production.

In fig. 27 the beam ellipses for different primary beam dimensions for 1 mm and 15 mm tungsten target are shown for photo-production (left plots) and electro-production (right plots). The energy of the positron beams is the one at which the maximum of production occurs.

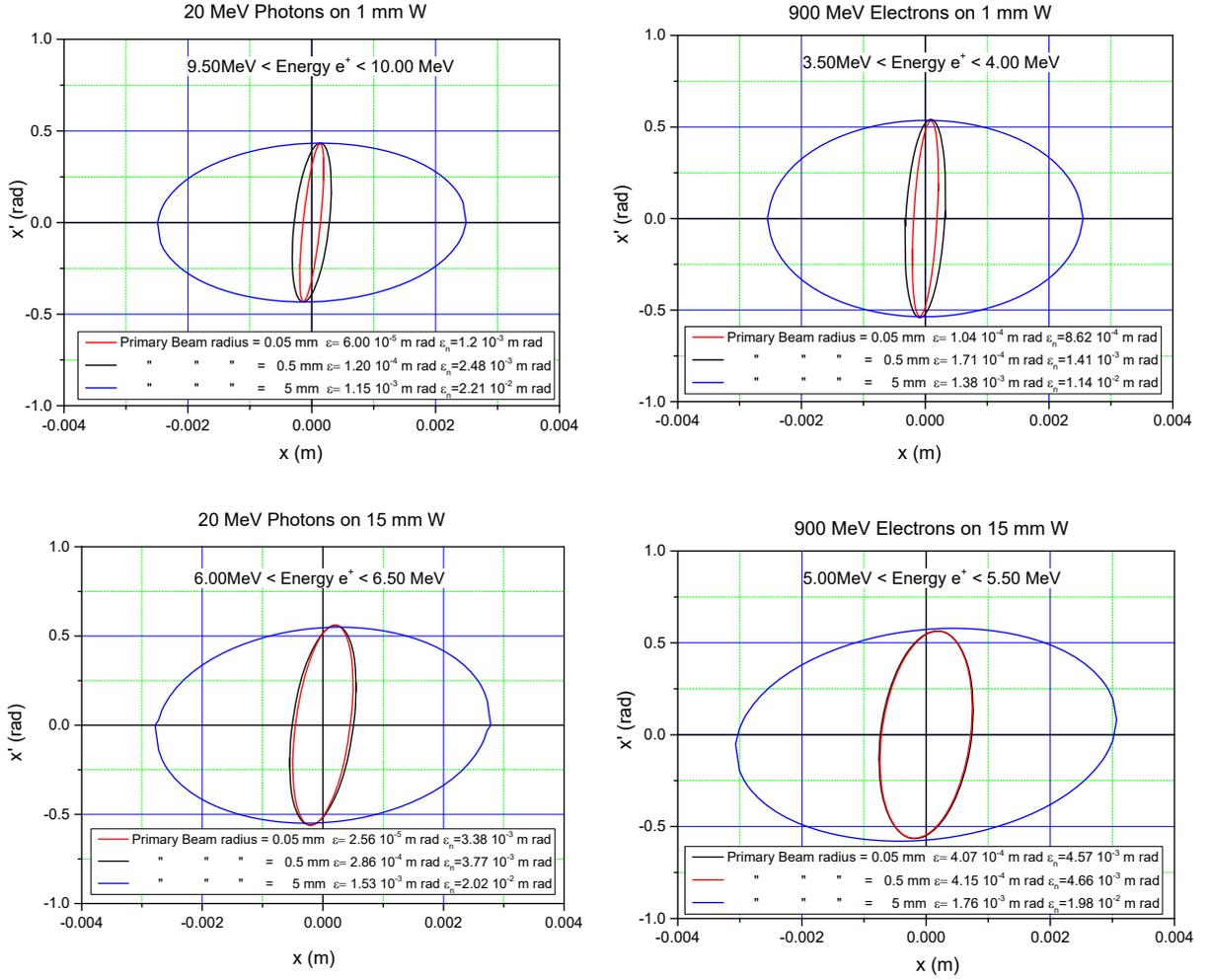

Figure 27: Positron beam ellipses at the energy where about the maximum of the production occurs for the different radii of the primary beams for 1 mm tungsten target (top plots) and for 15 mm tungsten target (bottom plots), for photo-production (left) and electro-production (right)

For thin target (top plots in fig.27) the different primary beam dimension effect is distinguishable, while for thick target (lower plots of fig.27) it is evident only the difference with the largest beam.

### 4.4.2 Emittance "a la Gruber"

In case of beam with large momentum spread, Gruber proposed a correction on the expression (4) for the normalized emittance [21], i.e.

$$\varepsilon_{n_G} = \sqrt{\sqrt{\langle x^2 \rangle \langle \beta^2 \gamma^2 x'^2 \rangle} - \langle xx' \beta \gamma \rangle^2} \qquad (5)$$

Of course in case of monochromatic beam or small momentum spread the two definitions (4) and (5) coincide.

Let's consider the case of photo-production by a 20 MeV photon beam on 1mm and 2 mm tungsten target, if we consider all the produced positrons ($6.91 \times 10^{-2} \pm 0.04\%$, and $9.24 \times 10^{-2} \pm 0.05\%$) respectively the energy spectrum spans from 0 to about 18 MeV (see fig. 10).

By considering energy slices of 0.5 MeV the two definition coincides (the maximum discrepancy is about 3%, occurring at the lowest energy, where the statistical accuracy is poor).

If we consider the whole energy range, for 1 mm target, the normalized emittance calculated with the two methods is $2.73 \times 10^{-3} \pm 0.02$ % m rad (from (4)) and $2.43 \times 10^{-3} \pm 0.02$ % m rad (from (5)) (see fig.14); being the variation about 12%

For the case of 2 mm we have $3.60 \times 10^{-3} \pm 0.02\%$ m rad (from (4)) and $3.31 \times 10^{-3} \pm 0.02\%$ m rad (from (5)) (see Fig.15); being the variation about 9 %.

As we can see the two values are quite closer each other (taking into account the statistical error and the statistics nature of the calculation), but in our case the energy/momentum spread is very high and we do not think applicable or useful this formula. As a matter of fact a beam with such characteristics cannot be transported along a beam line, if an energy selection is not performed.

### 4.4.3 Brightness

In the previous section we have already dealt about the "age" of the positron, i.e. the time the positron lives inside the target, from its creation to the time of escaping from the target. This is a useful parameter to evaluate the positron beam brightness.

The brightness is defined as:

$$B = \frac{2I}{\pi^2 \varepsilon_{nx} \varepsilon_{ny}} \qquad (6)$$

Where I is the beam current $\varepsilon_{nx}$ and $\varepsilon_{ny}$ are the transverse normalized emittances.

It is clear that the beam brightness depends, through the value of I, on the time length of the bunch. In order to give an absolute value of the characteristics of the positron beam produced we will consider the "intrinsic absolute brightness" by considering, instead of the bunch length, the positron bunch lengthening due to the production process.

As a matter of fact the primary photo/electron beam is considered as a $\delta$ and the produced positron at the exit from the target has his own age as told before and illustrated in fig.25 and 26. If we assume that the positron emerges from the target with a time distribution of the corresponding age, we can introduce the following definition of intrinsic brightness (per primary particle):

$$B(E) = \frac{2e}{\pi^2 \varepsilon_{nx}(E) \varepsilon_{ny}(E)} \frac{N_{e^+}(E)}{\Delta t(E)} \qquad (7)$$

Where $N_{e+}$ is the number of positron produced per primary particle and $\Delta t$ is the time spread of the age of the positrons (the standard deviation of the positron age), because a primary beam produces a secondary positron that escape from the target about some $10^{-10}$ seconds after its birth with an uncertainty (time lengthening) of about $10^{-14}$ - $10^{-16}$ seconds.

The absolute brightness of a positron beam will then be the one corresponding to the primary bunch time length + the time error ($\Delta t$).

As a numerical example we can consider a primary bunch 1 ps long, the intrinsic time spread of $10^{-14}$ - $10^{-16}$ is 0.02% − 2%, therefore we can conclude that the emitted positron beam has the same time structure of the primary beam for "long" bunches down to the ps scale. In fig. 28 brightness as from (7) is plotted for minimum (top plots) and maximum (lower plots) primary energy for photo-production (left) and electro-production (right) are shown for the various thickness. The "oscillations" at high energies in case of electro-production are due to the high statistical error.

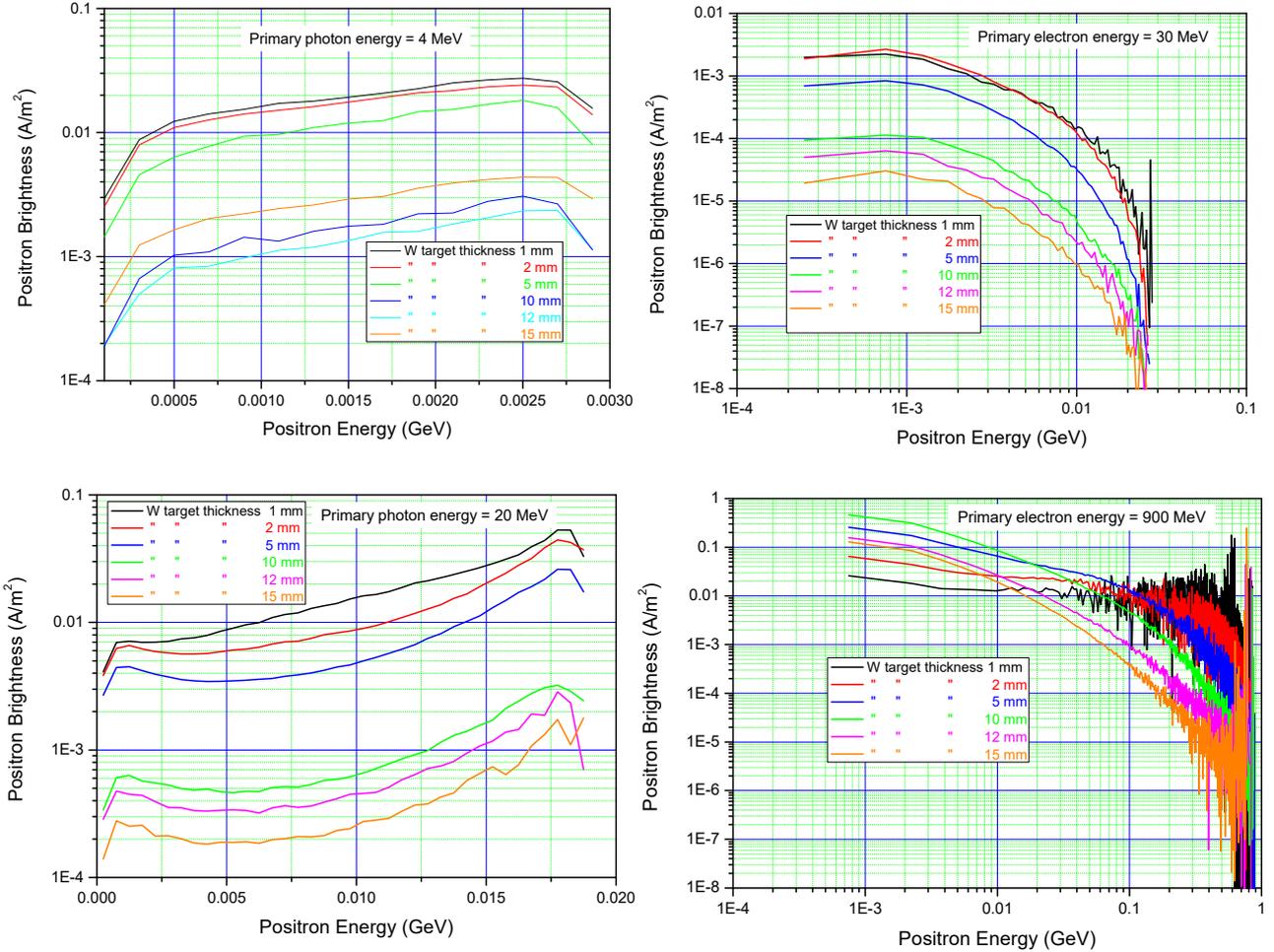

Figure 28: Positron beam brightness for 4 photo-production (left) and electro-production (right) for the minimum primary energy (top) and maximum primary energy (bottom) for different target thickness. At higher energy the values have high statistical error and big "oscillations" occurs.

From the plots we can see that in case of photo-production there is a correlation between brightness and age of the positrons, the decreasing of the brightness with the thickness is determined by the increasing of the age of the positrons. In case of electro-production the correlation is less evident in case of low energy primary (the curves relative to 1 and 2 mm are almost the same); for high energy electro-production there is no correlation between age/brightness and thickness.

## 5 ENERGY DEPOSITION IN THE TARGET

When the primary beam hits the target some energy is deposed and pressure waves [22] are created in the material; for high intensity or high repetition rates of the primary beam, these lead to damage of the target and the increase of the temperature can even cause the fusion of the target.

In the following table the energy deposition in the target is reported for the reference cases already examined.

Table 6 Energy deposition in the target for different primary radii

| W Target Thickness = 1 mm | | | |
|---|---|---|---|
| | | Photons ($E_\gamma$ = 20 MeV) | Electrons ($E_e$ = 900 MeV) |
| Primary Beam radius (cm) | Target Volume (cm$^3$) | Energy density ± err% (GeV/pr) | Energy density ± err% (GeV/pr) |
| 0.005 | 1.26E-02 | 3.58E-04 ± 0.11 | 2.73E-03 ± 0.04 |
| 0.05 | 1.26E-02 | 3.55E-04 ± 0.03 | 2.73E-03 ± 0.01 |
| 0.5 | 0.314 | 3.63E-04 ± 0.18 | 2.74E-03 ± 0.04 |
| W Target Thickness = 15 mm | | | |
| 0.005 | 0.189 | 1.06E-02 ± 0.09 | 0.25 ± 0.10 |
| 0.05 | 0.189 | 1.05E-02 ± 0.02 | 0.24 ± 0.04 |
| 0.5 | 18.85 | 1.33E-02 ± 0.06 | 0.32 ± 0.09 |

From the data we can see that the energy deposition is the same in case of thin target (1 mm), whatever the primary beam dimension is.

In case of thick target (15 mm) the energy deposition is about 30% higher for the large beam, while it is the same for the other two smaller radii. This is probably due to the spatial developing of the shower and the consequent energy deposition inside the target.

A very simple and conservative evaluation of the temperature increase can be done under adiabatic (very conservative) assumption.

Let's consider a primary beam of 900 MeV with a beam spot of 1 mm diameter hitting a 10 mm tungsten target (this the condition of maximum yield of 2.27/positron/primary production). In this case the energy deposition is about 0.133 GeV/primary electron. To produce $10^{12}$ positron/pulse, a primary intensity of $4.4 \times 10^{11}$ electron/pulse is needed.

If we assume a total deposition of the 900 MeV case by $4.4 \times 10^{11}$ electron primaries (about 10 J) in the volume given by the 1mm beam spot and 10 mm tungsten thick we obtain an increase in temperature of about 500 K. We must take into account the repetition rate of the pulses, and the diffusivity of the target, but so far it is impossible to guarantee the safe operation of the target.

Another estimation of the temperature increase has been given in [20].

The temperature increase is estimated by

$$\Delta T = 2N\eta \frac{E_{dep}}{cA} \tag{8}$$

Where $2N\eta$ is the amount of $e^+e^-$ pair emerging from the target

$E_{dep} \approx 2 \text{ MeV cm}^2 / g$ is the Energy deposition in the target assumed dominated by electron ionization loss

$c$ is the specific heat of the target (130 J/kg K for tungsten)

$A$ is the area of the beam spot

According to (8) the temperature increase in our case is about 280 K.

Again we must take into account the repetition rate of the pulses, and the diffusivity of the target, so a detailed thermal analysis is necessary, and this will be the argument of future work.

## 6  POLARIZATION

Polarization has not been so far considered, but the transfer of the primary polarization to the emerging positron according to [23] can be evaluated. This item will be developed in a future work.

## 7  CONCLUSIONS

The efficiency of the photo-production and electro-production has been evaluated.

The best target material is Uranium, being the highest Z element; anyway Tungsten (more easily manageable) gives about the same results.

Depending on the primary beam energy and target thickness the positron yield spans from 1.5e-3 $e^+$/pr to 2.27 $e^+$/pr, of which 1.7% to 45% is peaked in the forward direction (10° cone aperture).

The emittance and brightness of the produced positrons has been evaluated, showing a strong dependence on the primary beam transverse dimensions and time structure. The best choice between electro-production and photo-production related to the emittance and brightness of the positron beam produced depends on the energy of the primary beam and on the thickness of the target.

The time structure is preserved for pulses longer than some ps, being the intrinsic lengthening of about $10^{-14}$ - $10^{-16}$ seconds (1-2% of the bunch length for ps pulses).

The polarization has not been evaluated yet and will be investigated in the future.

In case of high intensity or repetition rate of the primary beam a thermal analysis is mandatory to evaluate the possible damage, up to the melting, of the target.

The polarization has not been evaluated yet and will be investigated in the future.

Secondary neutrons are emitted too; in case of experimental use of the produced positrons shielding may be necessary.